\documentclass[journal]{IEEEtran}
\usepackage{enumerate}
\usepackage{enumitem}
\usepackage{cite}
\usepackage{makecell}
\usepackage{booktabs}
\usepackage{graphicx}
\usepackage{array}
\usepackage{longtable}
\usepackage{multirow}
\usepackage{color}
\usepackage[colorlinks=true, allcolors=blue]{hyperref}
\usepackage{tabularx}
\usepackage{amsmath}
\usepackage{array}
\usepackage{pifont}
\usepackage{wasysym}
\usepackage{ragged2e}

\hypersetup{
colorlinks=true,
linkcolor=blue,
anchorcolor=blue,
citecolor=blue}

\renewcommand{\arraystretch}{1.4}

\newcolumntype{M}[1]{>{\centering\arraybackslash}m{#1}} 
\newcolumntype{J}[1]{>{\raggedright\arraybackslash}m{#1}}

\ifCLASSINFOpdf

\else

\fi

\begin{document}

\title{Agentic Satellite-Augmented Low-Altitude Economy and Terrestrial Networks: A Survey on Generative Approaches}

\author{Xiaozheng Gao, \textit{Member, IEEE}, Yichen Wang, Bosen Liu, Xiao Zhou, Ruichen Zhang, Jiacheng Wang, \\Dusit Niyato, \textit{Fellow, IEEE}, Dong In Kim, \textit{Fellow, IEEE}, Abbas Jamalipour, \textit{Fellow, IEEE},\\ Chau Yuen, \textit{Fellow, IEEE}, Jianping An, \textit{Senior Member, IEEE}, and Kai Yang, \textit{Member, IEEE} 

\thanks{Xiaozheng Gao, Yichen Wang, Bosen Liu, Xiao Zhou, and Kai Yang are with the School of Information and Electronics, Beijing Institute of Technology, Beijing 100081, China (e-mail: gaoxiaozheng@bit.edu.cn; wangyichen@bit.edu.cn; liubosen@bit.edu.cn; zhouxiao@bit.edu.cn; yangkai@ieee.org).} 
\thanks{Ruichen Zhang, Jiacheng Wang, and Dusit Niyato are with the College of Computing and Data Science, Nanyang Technological University, Singapore 639798 (e-mails: ruichen.zhang@ntu.edu.sg; jiacheng.wang@ntu.edu.sg; dniyato@ntu.edu.sg).}
\thanks{Dong In Kim is with the Department of Electrical and Computer Engineering, Sungkyunkwan University, Suwon 16419, South Korea (e-mail: dongin@skku.edu).}
\thanks{Abbas Jamalipour is with the School of Electrical and Computer Engineering, University of Sydney, Australia (e-mail: a.jamalipour@ieee.org).}
\thanks{Chau Yuen is with the School of Electrical and Electronics Engineering, Nanyang Technological University, Singapore 639798 (e-mail: chau.yuen@ntu.edu.sg).}
\thanks{Jianping An is with the School of Cyberspace Science and Technology, Beijing Institute of Technology, Beijing 100081, China (e-mail: an@bit.edu.cn).} 
}

\maketitle

\begin{abstract}
The development of satellite-augmented low-altitude economy and terrestrial networks (SLAETNs) demands intelligent and autonomous systems that can operate reliably across heterogeneous, dynamic, and mission-critical environments. To address these challenges, this survey focuses on enabling agentic artificial intelligence (AI), that is, artificial agents capable of perceiving, reasoning, and acting, through generative AI (GAI) and large language models (LLMs). We begin by introducing the architecture and characteristics of SLAETNs, and analyzing the challenges that arise in integrating satellite, aerial, and terrestrial components. Then, we present a model-driven foundation by systematically reviewing five major categories of generative models: variational autoencoders (VAEs), generative adversarial networks (GANs), generative diffusion models (GDMs), transformer-based models (TBMs), and LLMs. Moreover, we provide a comparative analysis to highlight their generative mechanisms, capabilities, and deployment trade-offs within SLAETNs. Building on this foundation, we examine how these models empower agentic functions across three domains: communication enhancement, security and privacy protection, and intelligent satellite tasks. Finally, we outline key future directions for building scalable, adaptive, and trustworthy generative agents in SLAETNs. This survey aims to provide a unified understanding and actionable reference for advancing agentic AI in next-generation integrated networks.
\end{abstract}


\IEEEpeerreviewmaketitle

\section{Introduction}\label{SectionI}
\IEEEPARstart{T}{he} emergence of the low-altitude economy networks (LAENets) \cite{d1_Unauthorized2025Li}, driven by the rapid advancement of unmanned aerial vehicles (UAVs) \cite{e10_LEO-Satellite-Assisted2021Jia}, urban air mobility (UAM) \cite{e5_The2025Shahriar}, and aerial logistics \cite{e4_Transfer2025Wen}, has introduced a new paradigm in digital infrastructure and intelligent services. These applications demand robust and ubiquitous connectivity across low-altitude domains \cite{e8_The2024Huang}, which traditional terrestrial networks alone cannot reliably support due to limited coverage, dynamic mobility, and challenging propagation conditions. In response, satellite-augmented low-altitude economy and terrestrial networks (SLAETNs) emerge as a critical infrastructure solution. The SLAETN integrates satellites (especially low-Earth-orbit (LEO) constellations \cite{e6_Low-Earth2024Lagunas}), LAENets, and terrestrial systems to deliver resilient and wide-area coverage for latency-sensitive low-altitude economy (LAE) applications \cite{e12_Reliable2022Liu}. This convergence aligns with the 6G’s vision of space-air-ground integrated networks (SAGINs) to support seamless, intelligent, and adaptive applications~\cite{r2_Near2024Liu}.

The integration of satellite, LAE, and terrestrial networks has attracted significant economic investment and policy recognition worldwide, underscoring its strategic value. 
Economically, according to the Civil Aviation Administration of China, China's low-altitude market is estimated to expand from $500$ billion Chinese yuan ($\approx70$ billion US dollars) in 2023 to $3.5$ trillion Chinese yuan ($\approx480$ billion US dollars) in 2035 \cite{e1_Secure2025Cai}.
Given connectivity's role as the operational backbone for scalable LAE services \cite{e3_A2021Baltaci}, satellite augmentation is quite essential. 
Studies confirm that LEO satellites provide a cost-effective solution for long-range connectivity while optimizing the coverage-latency trade-off \cite{e11_A2021Centenaro,e14_Performance2025Wang}. Notably, SpaceX's Starlink maintains over $1,000$ satellites at $550\,\text{km}$ altitude, while OneWeb operates more than $100$ satellites at $1,200\,\text{km}$ \cite{e2_A2025Le}, with both planning further deployments for global coverage.
Moreover, policymakers prioritize this convergence globally. In particular, China amends its Civil Aviation Law to support the LAE \cite{e15_2025China}, while Federal Aviation Administration's beyond visual line-of-sight (BVLOS) \cite{e16_A2024Betti} regulations accelerate LAE's commercialization in the United States. In parallel, the European Commission's Single European Sky Air Research program identifies satellite systems, e.g., Iris, as key enablers for next-generation air traffic management \cite{e17_Towards2024Gruber}. These initiatives demonstrate SLAETNs' critical role in next generation's economic growth.

Despite the promising benefits of SLAETNs, several critical challenges hinder their practical deployment and large-scale adoption. Specifically, these networks must navigate spatiotemporal unpredictability, platform heterogeneity, and mission-critical coordination across space-air-ground segments \cite{e18_Dynamic2023Zhang}. In such environments, the LAE applications that rely on real-time cross-segment collaboration, such as emergency response missions, cooperative UAV sensing, and high-mobility logistics, are especially vulnerable to disruptions caused by dynamic link failures, inconsistent protocol semantics, and uneven resource distributions \cite{e19_Hierarchical2022Chen}. 
Traditional solutions, such as rule-based strategies and static machine learning (ML) techniques, struggle under dynamic and data-sparse conditions due to their limited adaptability, semantic generalization, and real-time scalability \cite{s5_Deep2022De}.  Such constraints impede cross-layer coordination and intelligent scheduling, particularly in SLAETNs characterized by environmental dynamics, node heterogeneity, resource imbalance, and demand diversity. Consequently, these shortcomings ultimately impair SLAETNs to deliver high-quality, responsive, and autonomous  LAE services. 

To address these gaps, a new class of intelligent models known as agentic artificial intelligence (AI) \cite{u11_Internet2025Wang} has emerged. These models refer to autonomous intelligent agents with integrated perception-reasoning-action capabilities, capable of making multi-stage decisions and invoking tools based on real-time context \cite{u1_From2025Kshetri}. 
Unlike traditional task-specific models, agentic AI, inspired by human-like decision loops such as observe–orient–decide–act (OODA) \cite{u0_Parallel2023Yang}, emphasizes the integration of multi-stage cognition and interaction.
By integrating generative tools, such as real-time perception synthesis, environmental modeling, and policy generation, agentic AI systems demonstrate a greater degree of contextual awareness, goal-directed planning, and flexible interaction with uncertain and multi-domain network scenarios \cite{u3_Agentic2025Acharya}. In SLAETNs, these characteristics are crucial to managing highly dynamic aerial platforms, orchestrating cross-layer policies, and responding intelligently to unexpected mission-critical events. Therefore, such agents are particularly suitable for the complex task-driven environments within the SLAETN architecture. 

Within this paradigm, generative approaches, including generative AI (GAI) \cite{s7_Generative2024Liu} and large language models (LLMs) \cite{e21_Large2025Quan}, serve as crucial techniques, empowering agentic AI to model and simulate task environments. These generative techniques helps agentic AI to generate perceptual information, analyze scenarios, and formulate control strategies under uncertainty, thereby enabling agentic SLAETNs with situational awareness and autonomous behavior capabilities.
Specifically, GAI models refer to a class of models that can generate data, features, or control policies by learning latent distributions from training samples \cite{e20_An2023Park}. These models have demonstrated strong capabilities in synthesizing wireless propagation conditions, predicting mobility trajectories, and optimizing resource configurations in dynamic networks \cite{e22_Generative2025Betalo}. Meanwhile, LLMs provide semantic-level reasoning and abstraction across network layers, supporting natural language configuration, anomaly understanding, and policy synthesis \cite{e23_Integration2025Chen}. 
Together, GAI and LLMs form the cognitive and generative core of agentic AI systems, enabling a shift from reactive decision-making to predictive and autonomous behavior. The synergy between GAI and LLM provides adaptive perception, intelligent reasoning, and timely decision-making capabilities for the 6G networking \cite{f2_At2024Celik}, which can be further extended to a wider range of SAGINs or pure terrestrial/non-terrestrial networks (TNs/NTNs).

However, the integration of agentic AI into SLAETNs remains in  its early stage. In response to the evolving demands of LAE and recent advances in agentic AI, this survey provides a comprehensive guide for the field. It explores how the challenging issues in SLAETNs, such as real-time decision-making, resource-constrained sensing, and secure cross-domain coordination, can be addressed using agentic AI empowered by generative models, including GAI and LLMs. 
Table~\ref{tab:issues and solutions} presents a systematic structure to show the existing efforts on GAI and LLMs for the SLAETNs, from the perspectives of network communications, network security and privacy protection, as well as intelligent satellite tasks.

\begin{table}[ht]
\caption{Issues and Solutions \\ Red Circles Describe The Network Issues; Green Circles Represent The Overall Countermeasures For The Corresponding Issues; Green Check Markers Indicate Different Solutions Under Each Countermeasure}
\begin{center}
    \begin{tabular}{|llll|}
\hline
\multicolumn{4}{|l|}{Section III: Network communications} \\ \hline
\multicolumn{1}{|l|}{Issues}              & \multicolumn{3}{l|}{\begin{tabular}[c]{@{}l@{}}\textcolor{red}{\CIRCLE}~Dynamic channel environment~\cite{a11_Constructing2025Hao}\\ \textcolor{red}{\CIRCLE}~Limited information obtained~\cite{29_High-Altitude-UAV-Relayed2024Wang}\end{tabular}} \\ \hline
\multicolumn{1}{|l|}{Solutions}           & \multicolumn{3}{l|}{\begin{tabular}[c]{@{}l@{}}\textcolor{green}{\CIRCLE}~Channel modeling and estimation\\ \textcolor{green}{\ding{52}}~Channel modeling~\cite{a16_CWGAN-Based2024Li,1_Generative2024Zhang}\\ \textcolor{green}{\ding{52}}~Channel state information estimation~\cite{a14_Toward2023Machumilane,a41_Transformer2023Giuliano}\\ \textcolor{green}{\ding{52}}~Spectrum situation map construction~\cite{a17_Space-Based2024Yang,a2_SC-GAN2025Pan}\\ \textcolor{green}{\CIRCLE}~Network management and optimization\\ \textcolor{green}{\ding{52}}~Network decision making~\cite{3_Routing2025Guo,2_Generative2025Tao}\\ \textcolor{green}{\ding{52}}~Resource allocation~\cite{a3_Carrier2025Khoramnejad,5_Diffusion-Enabled2025Li,9_Generative2024Peiying}\\ \textcolor{green}{\ding{52}}~Network design and orchestration~\cite{a1_Generative2024Zhang,10_An2025Masoud,11_Space-air-ground2025Gao}\end{tabular}} \\ \hline

\multicolumn{4}{|l|}{Section IV: Network security and privacy protection} \\ \hline
\multicolumn{1}{|l|}{Issues}              & \multicolumn{3}{l|}{\begin{tabular}[c]{@{}l@{}}\textcolor{red}{\CIRCLE}~Confidentiality and covertness threat~\cite{b16_Multi-Objective2025Zhang,b33_Generative2024Liao}\\ \textcolor{red}{\CIRCLE}~Spoofing and intrusion attack~\cite{c29_Decentralized2018Milaat,c34_Distributed2020Li}\\ \textcolor{red}{\CIRCLE}~Cyber-physical attack~\cite{c15_Large2025Maatouk}\end{tabular}} \\ \hline
\multicolumn{1}{|l|}{Solutions}           & \multicolumn{3}{l|}{\begin{tabular}[c]{@{}l@{}}\textcolor{green}{\CIRCLE}~Confidentiality and covertness\\ \textcolor{green}{\ding{52}}~Secrecy energy efficiency optimization~\cite{4_Toward2025Kakati,24_Low-altitude2025Jia}\\ \textcolor{green}{\ding{52}}~Radio frequency fingerprint identification~\cite{a25_Radio2024Jiang}\\ \textcolor{green}{\ding{52}}~Covert communications~\cite{23_Covert2024Jia,a18_Res-GAN2024Wang}\\ \textcolor{green}{\CIRCLE}~Anti-spoofing and intrusion detection\\ \textcolor{green}{\ding{52}}~Anti-spoofing~\cite{a32_GNSS2021Li,a24_Machine2024Iqbal,a26_A2024Iqbal}\\ \textcolor{green}{\ding{52}}~Intrusion detection~\cite{a48_A2025He,a56_PLLM-CS2024Mohammed}\\ \textcolor{green}{\CIRCLE}~Adaptive security defense and signal recovery\\ \textcolor{green}{\ding{52}}~Security defense framework~\cite{8_Exploring2025Cao,31_An2025Qi}\\ \textcolor{green}{\ding{52}}~Signal recovery~\cite{21_Efficient2023Jing,22_High-precision2023Guo,a43_Diffusion2024Adam,25_GS3D2025Adam}\end{tabular}} \\ \hline

\multicolumn{4}{|l|}{Section V: Intelligent satellite tasks}   \\ \hline
\multicolumn{1}{|l|}{Issues}              & \multicolumn{3}{l|}{\begin{tabular}[c]{@{}l@{}}\textcolor{red}{\CIRCLE}~Dynamic orbital environment\\ \textcolor{red}{\CIRCLE}~Damaged sensing data quality\end{tabular}}   \\ \hline
\multicolumn{1}{|l|}{Solutions}           & \multicolumn{3}{l|}{\begin{tabular}[c]{@{}l@{}}\textcolor{green}{\CIRCLE}~Enhanced satellite network operations\\ \textcolor{green}{\ding{52}}~Satellite control and collision avoidance~\cite{a61_Language2024Victor,a39_LTE2024Jeong,a60_Intention2025Heng,a53_Enhanced2020Shen}\\ \textcolor{green}{\ding{52}}~Satellite transmission enhancement~\cite{a57_On-Air2024Hong-fu,a38_A2025Zhen}\\ \textcolor{green}{\CIRCLE}~Enhanced satellite sensing\\ \textcolor{green}{\ding{52}}~Climate prediction~\cite{a54_PrecipGAN2021Wang}\\ \textcolor{green}{\ding{52}}~Satellite-based positioning~\cite{a35_T-SPP2024Wu}\\ \textcolor{green}{\ding{52}}~Image processing~\cite{a13_MBGPIN2025Safarov,a46_Super-Resolved2025Ramirez-Jaime,a22_Free2025Chen}\end{tabular}} \\ \hline
\end{tabular}
\end{center}
\label{tab:issues and solutions}
\end{table}

\subsection{Related Surveys}

\subsubsection{Conventional AI for UAV and Satellite-Augmented Networks}
Recently, a growing body of surveys has explored conventional AI-driven approaches in NTNs, spanning UAV platforms, satellite communications, and their integration into space-air-ground architectures.
At the UAV platform level, the work in \cite{q1_Machine2024Kurunathan} offers a detailed investigation into ML techniques for UAV operations and communications. It categorizes learning-based methods into four core functional modules: perception, feature processing, trajectory planning, and flight control. The study further highlights the absence of unified and end-to-end ML pipelines for UAV automation.
Complementing this, the survey in \cite{q2_A2023Sai} provides a broader overview of AI in UAV systems, proposing a classification framework based on application domains, training paradigms, and algorithm types. Notably, it emphasizes the rise of reinforcement learning (RL) and federated learning (FL) as core drivers for enabling distributed and real-time UAV intelligence.

Shifting focus to satellite communications, the study in \cite{q7_Artificial2025Fontanesi} presents a comprehensive overview of AI/ML applications in satellite networks. It outlines use cases across different protocol layers and delves into the practical challenges of onboard implementation, including hardware constraints, radiation tolerance, and real-time processing. The survey also explores the potential of neuromorphic computing and commercial off-the-shelf components in future satellite AI systems.
In a similar domain, the authors in \cite{q6_Revolutionizing2024Mahboob} investigate satellite-based NTN architectures under the 6G vision, analyzing the role of AI in overcoming challenges such as beam management, Doppler shifts, and dynamic resource scheduling. Their work particularly highlights software-defined satellite operations as a key enabler for autonomous adaptability.

Beyond the satellite or UAV level, several surveys have turned attention to cross-layer integration in SAGINs. The work in \cite{q3_On2024Bakambekova} presents a system-level perspective on AI-empowered SAGINs, covering environment sensing, mobility control, and resource scheduling. In a more focused study \cite{q4_Resource2024Liang}, the authors reviews the resource allocation strategies in SAGINs, covering both optimization-based and learning-driven approaches. Finally, the study in \cite{q5_UAV-Assisted2024Arani} overviews the algorithmic solutions in UAV-assisted SAGINs, analyzing Q-learning, multi-armed bandits, and satisfaction-based learning under both 2D and 3D practical deployment scenarios.
Overall, the aforementioned surveys provide valuable insights into the adoption of conventional AI techniques across UAV-based, satellite-augmented, and their integrated network infrastructures. Consequently, these works lay the groundwork for intelligent perception, reasoning, and decision-making in SLAETNs.

\subsubsection{GAI/LLM for Networking}
The convergence of networking and generative intelligence has spurred a new wave of research interest \cite{r5_Generative2025Zhong}. Foundational surveys such as \cite{s3_Applications2025Vu} provide a comprehensive tutorial on how GAI techniques transform mobile and wireless networking through data synthesis, semantic modeling, and security enhancements. In \cite{s4_Deep2024Liu}, the focus narrows to deep generative models, such as generative diffusion models (GDMs), where a stepwise framework is proposed for their application in wireless network management. Complementing these, the study in \cite{s8_Generative2024Karapantelakis} offers a systematic taxonomy of GAI techniques tailored for mobile networks, covering performance requirements and standardization efforts by bodies like 3GPP. Furthermore, the authors in \cite{s11_Generative2025Khoramnejad} explores the role of GAI in optimizing 6G wireless networks, emphasizing scalable offline exploration and diverse scenario generation. Together, these works establish a technical foundation for integrating GAI with communication infrastructures.

Building upon this foundation, several studies zoom in on protocol-specific and task-specific GAI applications. For instance, the tutorial in \cite{s6_Enhancing2024Du} delves into GDMs applied to RL-based network optimization, supported by case studies in the specific scenarios such as semantic communications and vehicular networks. At the physical layer, the study in \cite{s9_Generative2024Van} discusses how generative approaches such as variational autoencoders (VAEs) and GDMs assist in channel modeling, signal classification, and blind equalization. Extending this perspective, \cite{s10_Generative2025Zhao} surveys the applications of GAI in enhancing physical layer security (PLS), addressing real-time confidentiality, availability, and integrity concerns. A specialized review in \cite{s12_Generative2025Liang} further investigates GAI-driven semantic communication networks by analyzing novel architectures, transceiver designs, and semantic knowledge management, with illustrative use cases in smart cities and Metaverse.

As GAI and LLMs continue to mature, their adoption in domain-specific and hierarchical networks has expanded. For example, the authors in \cite{s1_UAVs2025Tian} provides a forward-looking analysis of integrating LLMs with UAVs to enable agentic autonomy in perception, reasoning, and planning. Focusing on aerial network complexity, \cite{s13_Generative2024Nguyen} surveys GAI-driven optimization in aerial access networks (AANs), including UAVs, high-altitude platforms (HAPs), and LEO satellites. In a broader scope, the study in \cite{s2_Leveraging2025Javaid} discusses how LLMs can be embedded into SAGINs to support adaptive routing and intelligent resource management. In parallel, the study in \cite{s16_Unleashing2024Xu} investigates how AI-generated content (AIGC) services \cite{r4_Optimizing2025Lai} can be deployed at the mobile edge with privacy preservation and real-time customization. Nevertheless, a detailed analysis of generative approaches' role in agentic SLAETNs remains absent.

\subsubsection{Agentic AI for Autonomous Systems}
Agentic AI marks a shift from task-specific automation to autonomous and goal-driven intelligence. It goes beyond traditional AI agents by enabling systems to plan, learn, and collaborate with minimal human input across complex workflows. In \cite{u6_The2025Hosseini}, a broad narrative review focus on how agentic AI, characterized by autonomy, proactivity, and learning, transforms organizational operations through hierarchical agent architectures and tools such as LangChain, CrewAI, AutoGen, and AutoGPT. Also, it emphasizes the transition of model's position from “copilot” to “autopilot”, while addressing adoption challenges including risk, training, and ethical concerns. Complementing this, the authors in \cite{u9_Toward2025Zhang} present a conceptual taxonomy that distinguishes LLM-based AI agents from agentic AI systems. They map both design philosophies in terms of architecture, autonomy, interaction patterns, and task scope, demonstrating how agentic AI excels in memory persistence, tool integration, and collaborative problem. Together, these works define the conceptual foundation for building scalable agentic AI systems.

Building on these concepts, the practical implementation of agentic AI spans from industrial systems to communication networks. In \cite{u5_Industrial2025Mohammad}, a unified sense–infer–control (SIC) framework is introduced to enhance real-time monitoring and optimization across manufacturing, health, and motorsport domains. Generative and agentic models are shown to integrate with digital twins (DTs), enabling soft sensors and adaptive control. This idea is also extended into communication scope. In \cite{u7_From2025Jiang}, the authors develop a transition framework from large-AI-models (LAMs) to agentic AI for 6G, introducing modular components such as planners and knowledge bases to support multi-agent collaboration. The study in \cite{u8_AI2025Sapkota} further proposes a retrieval-augmented agentic framework for telecommunication networks. It improves network planning with structured knowledge and scenario reasoning. Additionally, the work in \cite{u10_Enhancing2025Laha} addresses hallucination issues in generative models through agentic workflows in 6G edge environments, thereby improving reliability and adaptability in dynamic settings.

As agentic AI systems grow toward more autonomous and collaborative, concerns surrounding their trustworthiness and safety become central. In \cite{u4_TRiSM2025Raza}, a trust, risk, and security management framework, termed as TRiSM, is introduced for multi-agent systems built on LLMs. This study identifies risks such as prompt manipulation and coordination failures and defines new metrics including component synergy score (CSS) and tool utilization efficacy (TUE) to evaluate the health of agentic AI systems. Strategies are proposed for privacy protection, explainability, and safe deployment. In the cloud domain, the study in \cite{u2_Agentic2025Patel} surveys agentic workflows for software-as-a-service (SAAS) support and troubleshooting, emphasizing domain-specific agents and LLM-powered reasoning with privacy preservation. Together, these works underscore the necessity of robust governance and security mechanisms for real-world deployment of agentic AI systems.

\subsection{Contributions of Our Survey}
Table~\ref{tab:related_surveys} summarizes existing surveys and tutorials across three major categories: conventional AI, generative techniques including GAI and LLMs, and the emerging field of agentic AI, within the context of intelligent systems.
While several existing surveys have begun to explore the use of GAI and LLMs in mobile systems, UAV platforms, or satellite-enabled integrated network architectures, a unified perspective on their role in enabling agentic AI within SLAETNs remains absent.
Distinct from prior surveys that either broadly discuss conventional AI applications in NTNs or narrowly focus on single GAI or LLM techniques in communication contexts, this work takes a different perspective. It specifically investigates how generative approaches including GAI and LLMs empower agentic AI across the space-air-ground layers of SLAETNs. The focus lies in three critical aspects: network communications, network security and privacy protection, and intelligent satellite tasks.
In particular, we highlight how GAI and LLMs contribute to key agentic capabilities, including perception generation, scenario construction, and policy synthesis, in task-driven and resource-constrained SLAETN scenarios.
By offering a generative approach-centric review from the perspective of agentic SLAETNs, this survey bridges a critical gap in the literature and lays a foundation for developing adaptive and autonomous systems in heterogeneous networks.

\begin{table*}[htbp]
\caption{Summary of Related Surveys on Conventional AI, Generative Techniques, and Agentic AI for Intelligent systems}
\centering
\begin{tabular}{p{3.8cm} p{1.5cm} p{6.2cm} p{4.9cm}}
\toprule
\textbf{Scope} & \textbf{References} & \textbf{Focus} & \textbf{Limitations} \\
\midrule
Conventional AI for UAV Networks & \cite{q1_Machine2024Kurunathan,q2_A2023Sai} & ML techniques for UAV control, perception, and trajectory planning; overview of AI in UAV system classfied by application domains and learning paradigms & Focus on conventional AI/ML; limited to UAV-specific context \\
Conventional AI for Satellite Networks & \cite{q7_Artificial2025Fontanesi,q6_Revolutionizing2024Mahboob} & Overview of AI in satellite protocols; discussion of onboard implementation, beam management, and software-defined-networking-enabled adaptability & Focus on conventional AI/ML; lack space-air-ground integrated context \\
Conventional AI for SAGIN Architectures & \cite{q3_On2024Bakambekova,q4_Resource2024Liang,q5_UAV-Assisted2024Arani} & AI-driven resource allocation, mobility control, and system-level coordination in SAGINs & Focus on conventional AI/ML; do not cover emerging generative models \\
GAI for General Wireless Networks & \cite{s3_Applications2025Vu,s4_Deep2024Liu,s8_Generative2024Karapantelakis,s11_Generative2025Khoramnejad} & Taxonomy of GAI models; GDMs for network tasks; standardization efforts & Mainly mobile-centric; lack space-air-ground integration context \\
GAI for Specific Network Functions & \cite{s6_Enhancing2024Du,s9_Generative2024Van,s10_Generative2025Zhao,s12_Generative2025Liang} & GAI for semantic communications, secure physical layer communications, and signal modeling; architecture and protocol-specific discussions & Limited cross-domain application view; lack agentic coordination focus \\
GAI/LLM for Domain-Specific and Hierarchical Networks & \cite{s1_UAVs2025Tian,s13_Generative2024Nguyen,s2_Leveraging2025Javaid,s16_Unleashing2024Xu} & GAI/LLM-driven decision-making, planning, and orchestration in aerial networks and SAGINs; edge deployment of AIGC services & Lack a unified generative-agent architecture and consistent model-task mapping across SLAETNs \\
Agentic AI: Foundations and Taxonomy & \cite{u6_The2025Hosseini,u9_Toward2025Zhang} & Distinctions between agentic AI and AI agents; strategic implications and organizational intelligence & Limited coverage of cross-layer implementation in communication networks \\
Agentic AI: System Architectures and Applications & \cite{u5_Industrial2025Mohammad,u7_From2025Jiang,u8_AI2025Sapkota,u10_Enhancing2025Laha} & SIC control for industry; LAM-to-agentic AI for 6G communications; agentic framework for telecommunication networks; agentic AI for edge computing & Lack fine-grained mapping to SLAETN segment-specific demand and optimization\\
Agentic AI: Trust, Risk, and Deployment Concerns & \cite{u4_TRiSM2025Raza,u2_Agentic2025Patel} & TRiSM for agentic systems; secure LLM workflows for cloud troubleshooting & Focus on enterprise/cloud systems; solely investigate agentic security management \\
\bottomrule
\end{tabular}
\label{tab:related_surveys}
\end{table*}

The key contributions of this paper are summarized as follows:
\begin{itemize}
  \item \textbf{Agentic Framework Perspective:} We introduce a novel agentic AI perspective that highlights how GAI and LLMs can enhance autonomous capabilities across perception generation, scenario construction, and decision synthesis for SLAETNs.
  
  \item \textbf{Model-Centric Comparative Analysis:} We systematically examine five classes of generative models, including VAEs, generative adversarial networks (GANs), GDMs, transformer-based models (TBMs), and LLMs, in terms of their unique generative mechanisms, practical applications, and performance trade-offs across heterogeneous network tasks.

  \item \textbf{Cross-Domain Application Mapping:} We present a structured analysis of how each generative model supports key tasks such as channel estimation, resource allocation, privacy enhancement, semantic control, and security defense across the SLAETNs.

  \item \textbf{Surveyed Literature Integration:} We comprehensively review more than $100$ key studies from both satellite-augmented networking and GAI communities, including classical seminal works and the latest research, to construct a unified and timely survey landscape.

  \item \textbf{Research Challenges and Future Directions:} We point out the limitations within existing studies through lessons learned for each network issue, such as model scalability, cross-domain transferability, and real-time adaptability. In addition, the future research directions for deploying agentic AI in dynamic SLAETNs are also discussed.
\end{itemize}

The remainder of this survey is structured as follows. Section~\ref{sec:background} introduces the system architecture of SLAETNs and outlines the roles of representative generative models. Section~\ref{sec:comm} discusses GAI/LLM-enhanced SLAETN communications, including channel modeling and estimation as well as network management and optimization. Section~\ref{sec:security} explores how generative techniques including GAI and LLMs reinforce security and privacy in SLAETNs, from different perspectives including communication confidentiality and covertness, anti-spoofing and intrusion detection, as well as adaptive security defense and signal recovery. Section~\ref{sec:application} examines diverse GAI/LLM-enhanced satellite tasks such as satellite control and sensing. Section~\ref{sec:challenges} outlines future research directions. Finally, Section~\ref{sec:conclusion} concludes the paper. Additionally, Table~\ref{tab:abbreviations} lists the abbreviations commonly used throughout this survey.

\begin{table*}[htbp]
\caption{List of abbreviations commonly used throughout this survey}
\centering
\begin{tabular}{llll}
\toprule
\textbf{Abbreviation} & \textbf{Description} & \textbf{Abbreviation} & \textbf{Description} \\
\midrule
AI & artificial intelligence & AIGC & artificial intelligence-generated content \\
ARC & autonomous reinforcement control & CDM & conditional diffusion model \\
BER & bit error rate & CNN & convolutional neural network \\
CoT & chain-of-thought & CSI & channel state information \\
CWGAN & conditional Wasserstein generative adversarial network & DD-GAN & data-driven generative adversarial network \\
DDPM & denoising diffusion probabilistic model & DDQN & double deep Q-network \\
DL & deep learning & DRL & deep reinforcement learning \\
DP-CDM & differentially private conditional diffusion model & DRA & diffusion-based resource allocation \\
DT & digital twin & FL & federated learning \\
GAI & generative artificial intelligence & GAIL & generative adversarial imitation learning \\
GAN & generative adversarial network & GDM & generative diffusion model \\
GNSS & global navigation satellite system & GS3D & Gaussian splatting-synergized stable diffusion \\
HAP & high-altitude platform & IID & independent and identically distributed \\
IoT & Internet of things & LAE & low-altitude economy \\
LAENet & low-altitude economy network & LAM & large-artificial intelligence-model \\
LEO & low-Earth-orbit & LLM & large language model \\
LLM-SA & large language model-based situation awareness & LoS & line-of-sight \\
LSTM & long short-term memory & MAIN & multi-agent intelligent networking \\
MAPPO & multi-agent proximal policy optimization & MBGPIN & multi-branch generative prior integration network \\
ML & machine learning & MSE & mean square error \\
MTS & multivariate time series & NLoS & non-line-of-sight \\
NTN & non-terrestrial network & PA & power amplifier \\
PCA & principal component analysis & PLS & physical layer security \\
PLLM & pre-trained large language model & PPO & proximal policy optimization \\
RAG & retrieval-augmented generation & ResGAN & residual GAN \\
RFFI & radio frequency fingerprint identification & RL & reinforcement learning \\
RMSE & root mean square error & RSS & received signal strength \\
SAC & soft-actor-critic & SAGIN & space-air-ground integrated network \\
SAR & synthetic aperture radar & SEE & secrecy energy efficiency \\
SIC & sense–infer–control & SLAETN & satellite-augmented LAE and terrestrial network \\
SNR & signal-to-noise ratio & SPP & single-point positioning \\
SR & secrecy rate & SSIM & structural similarity index measure \\
SSM & spectrum situation map & STIN & space–terrestrial integrated network \\
TBM & transformer-based model & TNN & transformer neural network \\
TVAE & tabular variational autoencoder & UAM & urban air mobility \\
UAV & unmanned aerial vehicle & UCB & upper confidence bound \\
VAE & variational autoencoder & VNE & virtual network embedding \\
\bottomrule
\end{tabular}
\label{tab:abbreviations}
\end{table*}

\section{Background Knowledge}~\label{sec:background}
In this section, we first introduce the SLAETNs from the definition, characteristics, and challenges. Subsequently, the potential of typical GAI models and LLMs to support agentic AI for specific applications in SLAETNs is discussed.

\subsection{Scenario Introduction: Satellite-Augmented LAE and Terrestrial Networks}
The SLAETN architecture represents a hierarchical and multi-domain framework that integrates three functionally distinct but collaboratively operated segments: LAENets, terrestrial networks, and satellite networks. This architecture leverages the flexibility of aerial mobility, the high-capacity service foundation of terrestrial infrastructures, and the wide-area coverage of satellite constellations. It enables a variety of emerging applications, such as urban aerial logistics, real-time environmental monitoring, emergency communication support, and global broadband connectivity. 

Each network segment of the SLAETNs serves a unique role and exhibits characteristics that introduce specific operational challenges, where the details are as follows.
\begin{itemize}
    \item \textbf{LAENets:}
    LAENets are composed of low-altitude platforms such as UAVs, UAM systems, and low-altitude autonomous sensing nodes. Operating primarily below $1000\,\text{m}$ and up to $3000\,\text{m}$ in special scenarios \cite{b1_Generative2025Chang}, they support diverse applications including aerial delivery, environmental inspection, and airborne relaying \cite{d2_Network2025Cheng}. Typically, these networks are characterized by highly dynamic 3D mobility, constrained onboard energy and computation, and intermittently connected links \cite{r1_Communications2025Yang}. As a result, LAENets face challenges such as complex trajectory planning, unreliable task offloading, and difficulty in executing real-time sensing and adaptation.

    \item \textbf{Terrestrial Networks:}
    Terrestrial networks form the core communication and computing substrate, consisting of base station infrastructures, edge/fog/cloud computing resources, and high-speed fiber backbones \cite{d13_From2025Ullah}. They provide high-throughput user access, real-time data processing, and orchestration of multi-type services. Key features include massive user access, heterogeneous service types, and a distributed multi-tier architecture \cite{d5_Machine2020Hussain,d6_Resource2019Abedin}. These characteristics give rise to challenges such as traffic congestion, load imbalance, stealthy traffic attacks, and inefficient task offloading across network tiers.

    \item \textbf{Satellite Networks:}
    Satellite networks, particularly those based on LEO constellations, serve as a critical enhancement layer by offering global coverage, seamless long-range communication, and resilient cross-domain coordination \cite{e13_Survey2016Radhakrishnan}. Besides extending services to remote or outage-prone areas, satellite networks provide high-precision positioning and stable line-of-sight (LoS) links to support aerial and terrestrial segments \cite{d11_Efficient2024Liu}. However, satellite systems are subject to long propagation delays, dynamic orbital topologies, partial topology observability, and open LoS links \cite{d12_Multi-Satellite2025Bakhsh}. Consequently, these factors lead to inefficiencies including feedback-based control, routing instability, topology-aware coordination difficulty, and vulnerability to link attacks.
\end{itemize}

Beyond the challenges inherent in each individual segment, the SLAETN architecture as a whole introduces additional cross-domain complexities that hinder system-level coordination. These include multi-layer protocol heterogeneity, spatiotemporal variability in connectivity, and limited global observability across network segments. Collectively, these factors present significant obstacles to unified service orchestration and end-to-end optimization, particularly under dynamic and uncertain conditions.

\subsection{Technique Introduction: GAI and LLMs in Satellite-Augmented LAE and Terrestrial Networks}
As SLAETNs evolve toward greater autonomy and responsiveness, there is a growing need for systems that not only sense and react, but also proactively reason, adapt, and make decisions under uncertainty. This calls for the adoption of agentic AI, which integrates perception, inference, and control in a goal-driven manner across dynamic and heterogeneous network environments. By incorporating generative techniques such as GAI and LLMs, agentic AI enables intelligent behavior that adapts to complex mission contexts and evolving link conditions \cite{d14_Large2025Guo}. Table~\ref{tab:challenges-solutions} presents the core characteristics and challenges of SLAETN components, along with GAI/LLM-driven solutions. Technically, GAI models support structured data synthesis, channel emulation, trajectory prediction, and policy generation under uncertainty \cite{b10_Emerging2024Kaleem}. In contrast, LLMs contribute semantic reasoning, multi-modal understanding, and instruction-following across network layers. Together, they form the cognitive engine of agentic AI, equipping SLAETNs with both low-level generative perception and high-level autonomous decision-making.

\begin{table*}[ht]
\renewcommand\arraystretch{1.2}
\belowrulesep=0pt
\aboverulesep=0pt
\centering
\caption{Network Characteristics, Challenges, and GAI/LLM-Driven Solutions}
\label{tab:challenges-solutions}
\begin{tabular}{>{\centering\arraybackslash}m{0.1\textwidth} || 
                >{\raggedright\arraybackslash}m{0.22\textwidth} | 
                >{\raggedright\arraybackslash}m{0.2\textwidth} | 
                >{\raggedright\arraybackslash}m{0.34\textwidth}}
\toprule
\midrule
\textbf{Network Type} & \textbf{Key Characteristics} & \textbf{Associated Challenges} & \textbf{GAI/LLM-Driven Solutions} \\
\toprule
\midrule

\multirow{3}{*}{
  \centering
  \begin{minipage}[c][1.4cm][c]{\linewidth}
    LAENets
  \end{minipage}
}
& High mobility of aerial nodes
& Difficult trajectory planning and coordination 
& \textbullet~GAI-based trajectory policy generation within RL framework\\ \cline{2-4} 

& Intermittent communication links
& Unstable communication and weak offloading reliability 
& \textbullet~GAI-based channel state information estimation \\ \cline{2-4}

& Limited onboard energy and computation resources
& Difficult to support real-time sensing and adaptation  
& \textbullet~LLM-based lightweight inference policy modeling \newline 
  \textbullet~GAI-based resource allocation strategy generation\\ \hline

\multirow{2}{*}{
  \centering
  \begin{minipage}[c][1.1cm][c]{\linewidth}
    Terrestrial\\Networks
  \end{minipage}
}
& Massive user access and heterogeneous services
& Traffic congestion and stealthy/mimic traffic attacks
& \textbullet~LLM-based intent-aware traffic modeling \\ \cline{2-4}

& Distributed multi-tier edge–cloud architecture
& Hard to orchestrate low-latency service offloading  
& \textbullet~LLM-based offloading policy generation \\ \hline

\multirow{4}{*}{
  \centering
  \begin{minipage}[c][3.6cm][c]{\linewidth}
    Satellite Networks
  \end{minipage}
}
& Long transmission distance and high latency
& Inefficient feedback and service delay
& \textbullet~GAI-based latency profile generation and content prediction \\ \cline{2-4}

& Highly dynamic orbital topology 
& Routing instability and handover complexity 
& \textbullet~GAI-based routing or handover strategy generation \\ \cline{2-4}

& Partial topology observability 
& Difficult to coordinate globally under uncertainty 
& \textbullet~LLM-based global topology modeling and task scheduling \\ \cline{2-4}

& Open line-of-sight links 
& Vulnerability to link attacks  
& \textbullet~GAI-based secure communication strategy generation \newline
   \textbullet~GAI-based spoofing/intrusion sample generation and detection \newline 
   \textbullet~LLM-based adaptive defense strategy generation \newline
   \textbullet~GAI-based signal recovery and image processing\\
\hline

\multirow{1}{*}{
  \centering
  \begin{minipage}[c][0.3cm][c]{\linewidth}
    SLAETNs
  \end{minipage}
}
& Multi-layer Heterogeneous Network Integration 
& Difficulties in end-to-end unified service orchestration and system-level optimization
& \textbullet~GAI/LLM-based protocol abstraction, inter-domain knowledge transfer, and coordination-aware policy generation \\ \hline

\end{tabular}
\end{table*}

Building on this foundation, the remainder of this subsection introduces representative GAI paradigms and LLMs, along with their respective generative capabilities in addressing SLAETN-specific tasks.

\begin{itemize}
    \item \textbf{VAEs: }
    By employing probabilistic latent space modeling, the VAE utilizes an encoder-decoder architecture to learn compressed representations through variational inference \cite{f1_Recurrent2018Wang}. This approach typically offers efficient uncertainty quantification and data reconstruction capabilities. 
    Within SLAETNs, these properties enable VAEs to support several practical tasks. Specifically, VAEs learn latent features of real channels to reconstruct missing or noisy channel state information (CSI) \cite{f3_How2023Saira}, helping to maintain stable communication under dynamic conditions. Besides, for physical-layer authentication, VAEs extract consistent RF signal traits to distinguish between devices \cite{f4_Physical-Layer2023Meng}. Moreover, when trained on legitimate data, VAEs identify spoofing data through reconstruction errors or latent outliers, enabling robust anti-spoofing detection \cite{f7_Dual2022Wu}. Also, VAEs can recover lost sensor signals under energy or bandwidth constraints \cite{f8_Cross-Modal2024Liang}, ensuring continuity in sensing and telemetry.
    To address more specialized demands, VAE variants improve their generative capabilities \cite{f6_eVAE2025Wu}. For example, conditional VAEs (CVAEs) \cite{f5_Physical2025Jiang} incorporate spatial or task-specific priors, enabling adaptive channel modeling and task-oriented signal generation. Tabular VAEs (TVAEs) \cite{f9_A2023Abraham} are tailored for structured data like routing tables or service schedules, thereby supporting intelligent orchestration across heterogeneous network segments.

    \item \textbf{GANs:}
    GANs employ an adversarial training mechanism between generator and discriminator networks to synthesize high-fidelity data \cite{f11_Recent2019Pan}. This framework excels in generating realistic samples under complex statistical constraints, which helps enhance the SLAETN performance. For example, in terms of spectrum situation map (SSM) reconstruction \cite{r12_Time2025Gao}, GANs generate high-fidelity radio maps from sparse measurements \cite{f13_RME-GAN2023Zhang}, which supports UAV path planning and spectrum coordination. When integrated with deep RL (DRL), GANs help simulate resource allocation strategies or represent network states under uncertain traffic or topology patterns, enhancing training efficiency \cite{f14_Experienced2021Kasgari}. Additionally, by generating adversarial examples to test and improve detection robustness \cite{b37_Robust2025Gaber}, they also play an efficient role in spoofing and jamming detections \cite{f12_Jamming2024Jiang}. Moreover, GANs also help reconstruct lost or corrupted waveform segments based on learned statistical distributions via adversarial training \cite{f15_ESR-GAN2021Kang}. Furthermore, by incorporating conditional constraints and Wasserstein distance, conditional Wasserstein GANs (CWGANs) \cite{f16_AVO2023Wang} improve stability and convergence in complex channel modeling tasks. Residual GANs (ResGANs) \cite{f10_Combining2022Deng}, which introduce skip connections in the generator, enhance detail preservation in SSM recovery and signal reconstruction, especially under noisy or incomplete input conditions. By utilizing real-world data, the generative adversarial imitation learning (GAIL) \cite{f17_Urban2024Wang} can develop expert-level strategies for resource allocation in hybrid networks.

    \item \textbf{GDMs:}
    GDMs synthesize data by progressively denoising samples drawn from a noise distribution via a learned reverse diffusion process \cite{f18_Diffusion2024Liu}. Their iterative and probabilistic formulation enables robust modeling of complex data, even under noise and perturbations.
    In SLAETNs, GDMs offer distinct advantages in multiple tasks. In terms of channel modeling and estimation, GDMs can capture nonlinear fading and dynamic patterns by simulating fine-grained channel realizations from limited measurements \cite{f22_Digital2025Gong}. Also, when combined with RL, GDMs can be trained to sample high-reward action sequences, thereby enhancing the convergence of RL-based scheduling \cite{b3_Diffusion-Based2025Zhang}. Therefore, they support efficient network decision-making and resource allocation in fast-varying topologies by exploring diverse strategies from complex solutions \cite{f23_Intention-Aware2025Liu}. Besides, GDMs learn normal traffic distributions and detect anomalous behaviors through reconstruction divergence, thereby enhancing the intrusion detection performance \cite{f21_Radio2023Zeng}. Moreover, to enhance signal recovery, GDMs can reconstruct clean waveforms by gradually filtering out interference, noise, or packet loss effects \cite{f19_CDDM2024}.  Furthermore, conditional diffusion models (CDMs) \cite{f20_Rate-Adaptive2025Yang} incorporate side information such as location, mobility state, or quality-of-service requirements, enabling scenario-specific channel synthesis and policy adaptation.  

    \item \textbf{TBMs:}
    Based on the self-attention mechanism \cite{f24_Attention2017Vaswani}, TBMs are generative architectures specifically designed to capture long-range dependencies and global structural patterns within sequential input data \cite{f24_A2024Ma}. This core capability, combined with their inherent scalability and parallelism, makes them exceptionally well-suited for high-dimensional modeling tasks that are prevalent in communication and sensing domains. Within the dynamic environments of SLAETNs, these properties enable TBMs to excel in several critical areas. Firstly, TBMs learn complex spatial-temporal attention patterns directly from sparse and noisy input sequences \cite{f25_Transformer-Based2024Zhang}, thereby providing accurate and real-time CSI predictions that are crucial for operations in high-mobility scenarios. Secondly, TBMs can generate attention-guided decision sequences that inherently model inter-user dependencies and adapt to time-varying channel dynamics \cite{f28_Transformer-based2025Han}, significantly enhancing coordination efficiency across heterogeneous network segments. Thirdly, by extracting cross-domain temporal features, TBMs effectively identify abnormal spectrum patterns and pinpoint spatial interference sources \cite{f27_GenAI-Based2024Saifaldawla}, enabling proactive mitigation strategies in congested orbital environments. Finally, TBMs can capture global visual dependencies to perform tasks including satellite image enhancement, denoising, and segmentation \cite{f26_A2025Wang}. This high-dimensional spatio-temporal analysis capability underpins their effectiveness across diverse network functions, such as environmental monitoring and disaster response.

    \item \textbf{LLMs:}
    Built upon large-scale pre-trained transformer architectures, LLMs demonstrate advanced capabilities in contextual reasoning, knowledge abstraction, and generative instruction following \cite{f29_A2024Hu}. Within SLAETNs, these capabilities enable transformative applications. Firstly, LLMs can parse natural-language operator commands into actionable instructions for network configuration \cite{f30_Survey2025Cao}, facilitating intuitive management and orchestration. Secondly, their semantic understanding supports the development of dynamic, semantic-aware security defense frameworks, automatically interpreting threats and generating mitigation policies \cite{f31_When2025Zhu}. Thirdly, LLMs enhance situational awareness by analyzing communications and telemetry for satellite intention recognition, which is critical for collision avoidance and cooperative operations \cite{f32_Unleashing2025Ni}. Crucially, LLMs' strong contextual generalization enables effective few-shot decision-making in complex scenarios involving protocol heterogeneity \cite{f33_AgentsCoMerge2025Hu}. These capabilities can significantly enhance the network resilience of satellite-augmented systems in dynamic, evolving environments that often have low observability.
    
\end{itemize}

\begin{figure*}[ht]
	\begin{center}
		\includegraphics[width=1\textwidth]{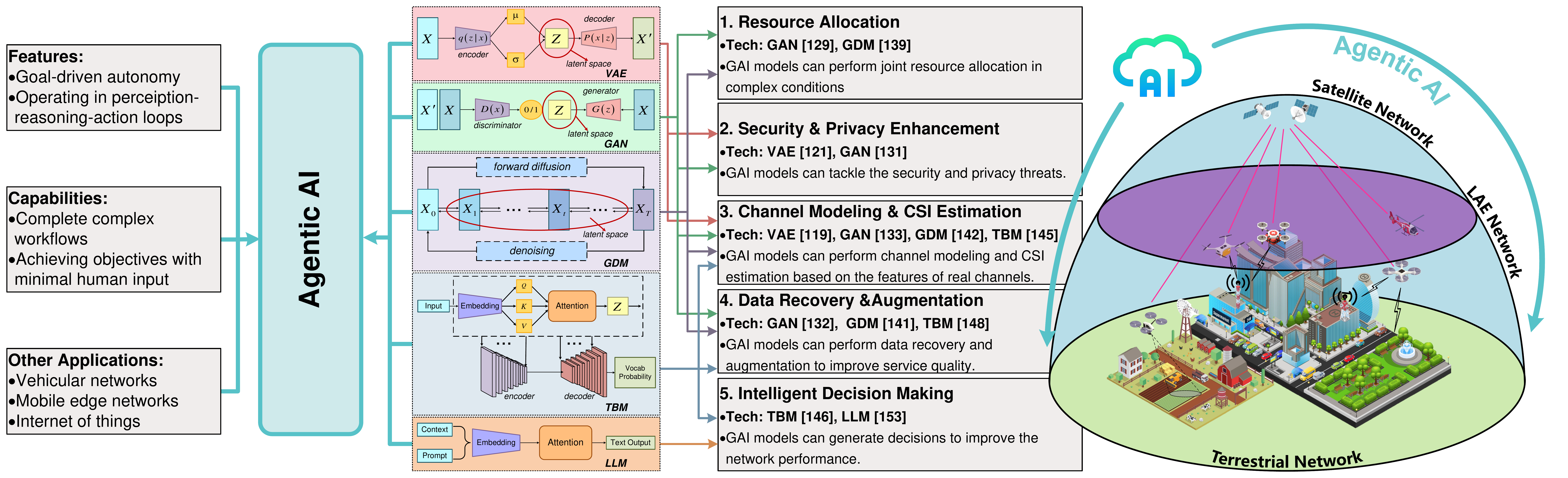}
		\caption{The overall architecture of the SLAETNs and the diagram of generative approaches for agentic AI. These GAI and LLM techniques are performed for diverse applications, such as resource allocation, security and privacy enhancement, channel modeling and CSI estimation, data recovery and augmentation, and intelligent decision-making.}
		\label{II}
	\end{center}
\end{figure*}

\textbf{Comparative Analysis:}
The generative techniques discussed above provide complementary capabilities essential for enabling perception, reasoning, and decision-making in agentic SLAETNs, as shown in Fig.~\ref{II}. Each paradigm of these generative techniques excels in specific scenarios based on inherent characteristics. 
Specifically, VAEs offer lightweight and uncertainty-aware inference, making them well-suited for anomaly detection, CSI recovery, and spoofing identification under constrained sensing conditions \cite{f36_Generative2025Deng}. Compared to VAE, GANs excel in synthesizing high-fidelity data samples and adversarial patterns, supporting spectrum map emulation, waveform recovery, and robust threat simulation \cite{f35_Generative2018Creswell}. Compared to VAEs and GANs, GDMs exhibit stronger robustness to noise due to their iterative generation capability, making them ideal for fine-grained decision sampling, semantic restoration, and proactive policy exploration in dynamic environments \cite{f37_A2024Cao}.
In contrast, TBMs are particularly effective for spatiotemporal pattern extraction and long-sequence modeling, which are crucial for mobility-aware scheduling, interference source tracking, and satellite image understanding \cite{f38_A2022Dai}. Finally, as high-level semantic agents, LLMs provide powerful cross-domain reasoning and abstraction capabilities. They are uniquely suitable for translating natural language instructions into structured control policies, orchestrating multi-segment missions, and enabling human-centric autonomous interaction across heterogeneous infrastructures \cite{f34_Recent2024Hagos}.
In summary, a system should consider application properties and performance goals to select appropriate generative techniques for agentic network tasks. For example, VAEs are preferred for efficient uncertainty quantification, GANs for high-fidelity signal emulation and robustness training, GDMs for noise-resilient data generation and sequential decision making, TBMs for real-time temporal-spatial coordination, and LLMs for semantic orchestration and policy abstraction. A hybrid deployment of these paradigms offers a flexible and scalable path toward intelligent, adaptive, and autonomous network behaviors.

\section{GAI and LLMs for SLAETN Communication Enhancement}~\label{sec:comm}
Channel modeling and estimation as well as network management and optimization are fundamental for enabling reliable connectivity and efficient coordination in SLAETN communications. However, real-time tracking of highly dynamic channel states and scalable optimization under stringent latency constraints are significantly challenged \cite{a11_Constructing2025Hao,29_High-Altitude-UAV-Relayed2024Wang}. This section provides an overview of the applications of GAI and LLMs for agentic SLAETN communication enhancement.

\subsection{Channel Modeling and Estimation}

Channel modeling and estimation in SLAETNs constitute a critical foundation for ensuring reliability in space-air-ground cross-domain communications. 
Therefore, it is imperative to precisely model the dynamic and heterogeneous channels within SLAETN environments. However, conventional model-driven methods often fail to capture the intricate channel characteristics \cite{b35_Generative2022Xia,r11_Ultra2025Hua}, while conventional ML methods, such as convolutional and recurrent neural networks (CNNs/RNNs), are always constrained by limited generalization capabilities and computational inefficiency in real-time scenarios \cite{b22_Machine-Learning-Based2022Mao}. 
In this context, GAI methods offer a novel data-driven solution by capturing spatial-temporal variations \cite{b42_Spatially2024Giuliani}, generating high-fidelity synthetic data \cite{b28_LSTM-Based2023Rasheed}, and estimating CSI from sparse or outdated observations \cite{b38_Distributed2022Zhang}. Additionally, GAI-enabled SSM construction exploits spatial correlations to provide valuable priors for propagation modeling and CSI inference \cite{b11_3D2023Hu}. These capabilities of GAI collectively enable accurate, adaptive, and real-time wireless channel characterization in complex SLAETNs.

\subsubsection{Channel Modeling}
In the context of agentic AI, accurate and adaptive perception of the wireless environment is foundational. GANs, characterized by the generator and discriminator components, enable direct learning of wireless channel probability distributions from empirical data without reliance on explicit mathematical modeling. However, traditional GANs, which employ Jensen-Shannon divergence to measure distribution discrepancies, are prone to inducing mode collapse and suffering from gradient vanishing issues \cite{c2_Generative2014Goodfellow}. 
Addressing the convergence issues caused by weight clipping, the authors in \cite{a16_CWGAN-Based2024Li} proposed a CWGAN with gradient penalty (CWGAN-GP)-based method for end-to-end channel modeling in satellite-terrestrial integrated networks (STINs). The CWGAN-GP generates realistic channel responses and enhances perceptual fidelity by learning conditional distributions of satellite-terrestrial channels under limited real data, with the Wasserstein framework \cite{c43_AVO2023Wang} improving controllability and the GP mechanism \cite{c44_A2024Roy} ensuring stable training. The simulations demonstrate that the proposed CWGAN-GP-based scheme achieves similar bit error rate (BER) and block error rate compared to benchmarks while exhibiting an advantage in terms of computational complexity. 

In parallel, GDMs further enrich the perceptual accuracy of agentic AI, by capturing nonlinear non-LoS (NLoS)-induced multipath dynamics through iterative denoising processes \cite{c3_Exploring2024Du}. This capability makes them well-suited for constructing channel maps that reflect realistic propagation conditions. Based on this, a GDM-based channel information map construction framework was introduced in \cite{1_Generative2024Zhang}. This framework leverages a pipeline of image acquisition, denoising, and zero-shot classification to extract LoS signal features from the noisy data, thereby generating accurate channel information maps for dynamic SAGIN environments. Notably, compared to traditional ML methods like CNNs, the GDM achieves higher accuracy, F1 score, and recall, while requiring significantly fewer iterations to converge, demonstrating its efficiency in extracting complex data structures.

\subsubsection{CSI Estimation}
Based on accurate channel modeling, obtaining the CSI, especially the LoS/NLoS state information of communication links \cite{r9_Three2022Meng}, is also crucial for the decision-making issues in SLAETNs, such as load balancing or traffic scheduling. However, although RL-based schedulers are emerging, the agents can only observe the status of selected links, resulting in extremely slow convergence \cite{c50_Learning-Based2023Machumilane}. Consequently, agents with partially observable states may fail to adapt to the very short satellite visibility periods of large constellations \cite{c51_Partially2007Jason}. To this end, the authors in \cite{a14_Toward2023Machumilane} adopted a conditional tabular GAN (CTGAN) for offline training to learn from large-scale prior data and generate CSI. Meanwhile, a TVAE was used for online learning to further fine-tune and optimize the process, ensuring continuous and accurate CSI provision. These generative models effectively enhance perception and real-time reasoning in agentic AI systems. The simulations confirm that the proposed methods significantly improve LoS estimation and traffic scheduling, reduce end-to-end loss, and balance offline accuracy with online efficiency.

Furthermore, to address the CSI aging issue caused by rapid dynamics in STIN, the authors in \cite{a41_Transformer2023Giuliano} proposed a transformer neural network (TNN)-based CSI prediction method. Specifically, the TNN is employed to predict the signal-to-interference-plus-noise ratio (SINR) of each user equipment by learning temporal dependencies from historical CSI reports. This generative prediction process demonstrates anticipatory reasoning of agentic AI, enabling accurate modulation and coding scheme selection without frequent CSI feedback, particularly effective under high-latency NTN conditions. The simulations show that compared to the conventional periodic CSI reporting schemes, the proposed method improves throughput by approximately $5\,\text{Mbps}$ with only $1.2\,\text{ms}$ prediction delay.

\subsubsection{SSM Construction}
The SSM directly reflects the signal propagation characteristics within electromagnetic environment, offering spatial priors for estimating channel parameters, such as path loss and shadow fading \cite{c4_Channel2021Hu}. Therefore, it serves as a valuable complement to conventional channel estimation, especially in satellite communication scenarios where accurate CSI is often difficult to obtain. Compared to ground-based monitoring, space-based monitoring faces unique challenges, including long data update cycles and the difficulty of detecting dynamically changing narrow-beam signals \cite{c5_SpectrumChain2023Wu, c6_Interference2021Hao}. To enable space-based SSM construction, a GAN-based approach enhanced with dilated convolutions and ResNet connections, known as DCRGAN, was proposed in \cite{a17_Space-Based2024Yang}. This model generates complete SSMs by learning the spatial characteristics of received signal strength (RSS) and inferring unmeasured regions. These generative capabilities demonstrate the perceptual intelligence of agentic AI that infers complete environmental representations from limited and noisy data. The integration of dilated convolution \cite{c45_Spatiotemporal2022Ma} and residual learning \cite{c46_A2023Jiang} strengthens contextual feature representation and contributes to stable and accurate generation performance, thereby supporting robust feature abstraction and spatial reasoning. The simulations demonstrate that the proposed DCRGAN outperforms the baselines in the SSM construction performance comparison in terms of average mean squared error (MSE) and computational efficiency. 

\begin{figure}[!t]
	\begin{center}
		\includegraphics[width=1\columnwidth]{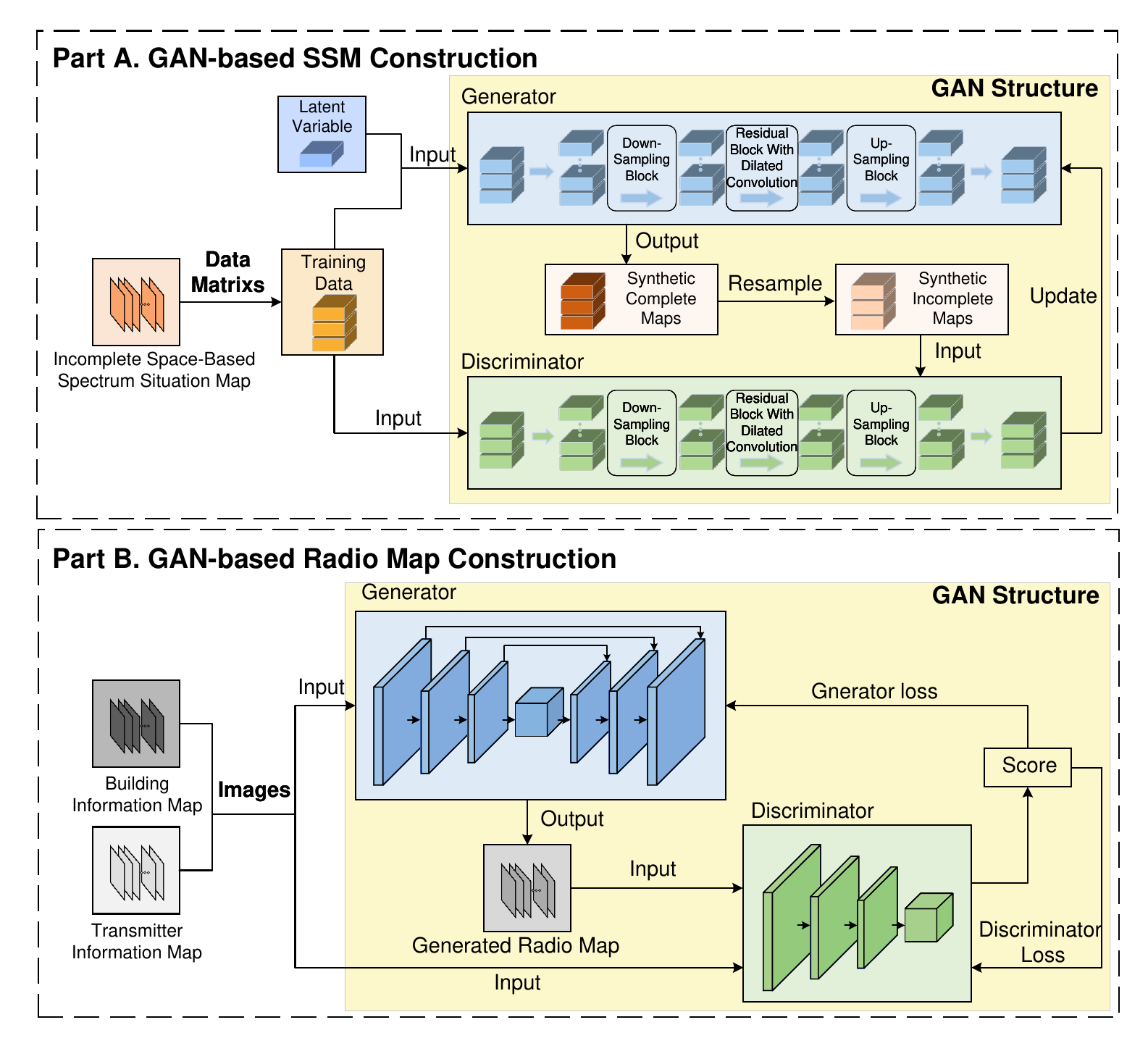}
		\caption{This illustration presents the GAN-based SSM construction methods. Part A illustrates the GAN-based SSM construction scheme in \cite{a17_Space-Based2024Yang}, which learns the characteristics of received signal data. While Part B illustrates the GAN-based radio map construction scheme in \cite{a2_SC-GAN2025Pan}, where building outlines and transmitter information are directly utilized as the inputs.}
		\label{III-A-3}
	\end{center}
\end{figure}

Nevertheless, the DCRGAN proposed in \cite{a17_Space-Based2024Yang} depends on sparse RSS measurements, making the radio map quality sensitive to sample distribution, while also incurring high acquisition costs and being vulnerable to malicious interference. Addressing these limitations, a Pix2Pix GAN-based spectrum cartography algorithm was proposed in \cite{a2_SC-GAN2025Pan} to reconstruct high-resolution radio maps from satellite-acquired building images and transmitter information. In this framework, the generator learns a direct mapping from geographic scenes to signal strength distributions, while the discriminator refines the generation quality via patch-wise adversarial training. This architecture exemplifies context-aware generation, a key trait of agentic AI, wherein structural consistency and domain semantics are enforced during spectrum map synthesis. The experimental results show that the proposed method outperforms the benchmarks in terms of structural similarity index measure (SSIM) and root MSE (RMSE). Notably, as illustrated in Fig.~\ref{III-A-3}, these GAN-based SSM construction frameworks highlight how generative models enable agent AI's perception and reasoning, which is crucial for adaptive cognition in spatially distributed SLAETNs.

\textbf{Lesson Learned.} 
As summarized in Table~\ref{GAI in Channel Modeling and Estimation}, GAI has demonstrated its great potential in enhancing channel modeling and estimation in SLAETNs. This lies in its capabilities of capturing complex spatial-temporal features and enabling accurate reconstruction from limited or outdated data. However, current GAI-based models often suffer from high training overhead due to complexity. For example, the U-Net-based GDM proposed in \cite{1_Generative2024Zhang} contains approximately $8$ to $10$ million parameters and requires around $12$ to $15$ hours for training. Besides, the Pix2pix2 GAN in \cite{a2_SC-GAN2025Pan} comprises about $12$ million parameters and needs $5$ to $8$ hours of training time. Consequently, these limitations hinders the deployment in real-time and resource-constrained SLAETN environments. Future research should focus on developing lightweight models with real-time adaptability.

\begin{table*}[ht]
\renewcommand\arraystretch{1.2}
\belowrulesep=0pt
\aboverulesep=0pt
\centering
\caption{GAI in Channel Modeling and Estimation}
\label{GAI in Channel Modeling and Estimation}
\begin{tabular}{>{\centering\arraybackslash}m{0.11\textwidth} | | >{\centering\arraybackslash} m{0.03\textwidth} | >{\centering\arraybackslash}m{0.08\textwidth} | >{\raggedright\arraybackslash} m{0.21\textwidth} | >{\raggedright\arraybackslash}m{0.21\textwidth}| >{\raggedright\arraybackslash} m{0.21\textwidth}}
\toprule
\midrule
    Type & \multicolumn{1}{c|}{Ref} & Algorithm & Description & Pros & Cons \\ \toprule\midrule
\multirow{2}{*}{
  \centering
  \begin{minipage}[c][2.2cm][c]{\linewidth}
    Channel Modeling
  \end{minipage}
}      
    & \cite{a16_CWGAN-Based2024Li} & CWGAN-GP & CWGAN-GP is employed for end-to-end channel modeling in STIN communications. 
    & \textbullet \ Model-free adaptability to complex channels \newline  \textbullet~Improved training stability 
    & \textbullet~Dependence on sufficient training data \newline \textbullet~High computational overhead for training \\ \cline{2-6} 
    & \cite{1_Generative2024Zhang} & GDM & GDM is employed to construct channel information map for SAGINs. 
    & \textbullet \ Superior nonlinear modeling capability \newline  \textbullet~Robust noise suppression and feature extraction \newline \textbullet~Data-efficient zero-shot learning 
    & \textbullet~High computational overhead for training \newline \textbullet~Sensitivity to initial noise parameters \\ \hline
\multirow{2}{*}{
  \centering
  \begin{minipage}[c][1.6cm][c]{\linewidth}
    CSI Estimation
  \end{minipage}
}    
    & \cite{a14_Toward2023Machumilane} & CTGAN and TVAE  & CTGAN provides high-accuracy offline CSI generation, while TVAE enables adaptive online learning under small-sample constraints in NTN networks. 
    & \textbullet \ Effective handling of tabular data \newline \textbullet~Real-time training feasibility \newline
    \textbullet~Acceleration of RL convergence
    & \textbullet~Dependence on data size and correlation \newline \textbullet~Long training time for CTGAN \\ \cline{2-6} 
    & \cite{a41_Transformer2023Giuliano}& TNN & TNN is employed to address the CSI aging issue caused by rapid dynamics in STIN. 
    & \textbullet \ Superior handling of long sequences and gradient issues \newline \textbullet~Parallel Processing Efficiency 
    & \textbullet~Dependence on high-frequency training data \newline \textbullet~High computational overhead for training\\ \hline
\multirow{2}{*}{
  \centering
  \begin{minipage}[c][2.2cm][c]{\linewidth}
    SSM Construction
  \end{minipage}
}   
    & \cite{a17_Space-Based2024Yang} & DCRGAN & DCRGAN is employed for the space-based SSM construction from sparse RSS measurements. 
    & \textbullet \ High-fidelity SSM completion \newline \textbullet~Robust Training Framework 
    & \textbullet~Dependence on sparse RSS measurements \newline \textbullet~High computational overhead for training\\\cline{2-6} 
    & \cite{a2_SC-GAN2025Pan} & Pix2Pix GAN & Pix2Pix GAN is employed to reconstruct high-resolution radio maps from satellite-acquired data in STIN. 
    & \textbullet \ Effective capture of geometric relationships \newline \textbullet~Reduced dependence on sparse sampling \newline \textbullet~Enhanced adaptability to dynamic scenarios
    & \textbullet~Dependence on high-quality satellite data \newline \textbullet~High computational overhead for training \newline 
    \textbullet~Sensitivity to transmitter parameter variations \\ \hline
\end{tabular}
\end{table*}

\subsection{Network Management and Optimization}

In SLAETNs, effective network management and optimization is critical to ensure reliable, efficient, and seamless service delivery. However, traditional methods, primarily reliant on rule-based systems and static optimization algorithms, struggle to cope with the large scale, rapid topological changes, and complex inter-dependencies inherent in SLAETNs \cite{c7_Space-Air-Ground2018Liu,c8_Deep2024Huang}. Conversely, recent advances have demonstrated the significant potential of GAI and LLMs in enabling intelligent and data-driven network management and optimization, such as real-time network decision-making \cite{a44_Generative2024Wang}, dynamic resource allocation \cite{13_Generative2025Song,a20_Dynamic2024Li}, and intelligent network design and orchestration \cite{a19_Enhancing2024Sheikh}. 
In terms of network decision-making and resource allocation, GAI-based models, particularly GANs and GDMs, are utilized to tackle dynamic constraints and model high-dimensional resource features.
While in terms of network design and orchestration, LLM-based frameworks are employed to automate the complex coordination of diverse network segments \cite{30_Utilizing2024Tang}, by leveraging the global cognitive capabilities. Overall, GAI and LLMs empower SLAETN management with unprecedented adaptability, intelligence, and scalability.

\subsubsection{Network Decision-Making}
Addressing the real-time decision-making under dynamic topologies in SLAETNs  constitutes a fundamental basis for network resource allocation. However, traditional DRL always incurs high overhead, long convergence time, and latency due to its reliance on continuous real-time data, particularly inefficient in large-scale network decision making scenarios \cite{c9_Survey2021Tang}. To overcome these limitations, the authors in \cite{3_Routing2025Guo} developed a GAN-powered deep Q-network (DQN) framework to enhance adaptive routing under dynamic topologies and long-latency communications within SAGINs.  In this framework, the GAN learns the distribution of historical network states and synthesizes representative features to augment the DQN input. This generative capability serves as an enabler of reasoning and anticipatory decision-making in agentic AI systems, allowing robust routing policy learning with limited real-time input. The simulations show that the GAN-DQN scheme reduces average latency by approximately $20\%$ compared to pure DQN scheme while maintaining comparable packet loss rates under high-traffic conditions.

Moreover, by leveraging the forward noising and reverse denoising processes, GDMs hold the potential in capturing the highly dynamic and multidimensional environmental characteristics of SLAETNs, including satellite mobility and user distribution \cite{s5_Deep2022De}. These capabilities position GDMs as effective decision policy generators within agentic AI, capable of producing diverse and feasible responses under uncertainty \cite{c11_Diffusion-Based2024Du}. Based on this, the authors in \cite{2_Generative2025Tao} developed a diffusion model-assisted multi-agent proximal policy optimization (DMAPPO) framework for robust vertical handover decision-making in Internet of things (IoT)-based SAGIN scenarios. In this framework, the diffusion model acts as the policy network to generate diverse and feasible handover decisions, which are then refined through MAPPO to enhance generalization and convergence in dynamic environments. This process embodies multi-agent deliberation and adaptive strategy synthesis, critical to agentic AI's behavior in SLAETNs. The simulations demonstrate a reduction in handover failure rate by up to $58\%$, alongside improved Pareto efficiency compared to conventional MAPPO algorithms. 

\subsubsection{Resource Allocation}
Based on the aforementioned low-level decision-making tasks, it is critical to resolve the cross-domain resource allocation for enhancing resource utilization efficiency of SLAETNs \cite{r14_Dependency-Elimination2025Ouyang}. 
In response to the inefficiency of traditional RL methods, e.g. double DQN (DDQN) \cite{c47_Delay-Aware2022Khoramnejad} and upper confidence bound (UCB) \cite{c48_Improved2021Hashima}, in exploring high-dimensional dynamic environments, a denoising diffusion probabilistic model (DDPM)-enhanced DRL method was proposed in \cite{a3_Carrier2025Khoramnejad}. The proposed method aims to optimize the joint problem of carrier aggregation, load balancing, and backhauling (JCALB) in LEO satellites-based NTNs. Specifically, the DDPM enables diverse exploration by generating candidate JCALB strategies from complex solution spaces via its denoising mechanism \cite{c49_Conditional2025Letafati}, while DRL refines these strategies based on dynamic network feedback. This combination equips the system with agentic exploration and adaptive policy synthesis, allowing for robust decision-making under uncertainty. Impressively, compared to the DDQN- and UCB-based baselines, the proposed method enhances the transmission capacity of LEO satellites while utilizing fewer carriers and subchannels, achieving a $42\%-63\%$ reduction in load factors.

Similar to the DDPM-based method in \cite{a3_Carrier2025Khoramnejad}, a diffusion model-based resource allocation (DRA) algorithm was introduced in \cite{5_Diffusion-Enabled2025Li}, to support DT synchronized resource allocation in SAGINs. By iteratively denoising sampled trajectories from Gaussian noise, the DRA is employed to generate synchronized DT representations from sparse or delayed observations. This capability reflects perceptual reconstruction and environmental inference, two foundational features in agentic AI, enabling accurate downstream resource allocation with partial input. Compared to DRL and heuristic algorithms, the DRA scheme reduces time-average energy consumption by $14.3\%-26.8\%$ and transmission queue length by $19.56\%-34.37\%$, effectively improving the resource utilization.

Moreover, existing DRL-based virtual network embedding (VNE) algorithms often rely on manual reward design \cite{c13_Novel2018Cao}, struggling to address the inequitable resource allocation in SLAETNs. In this context, a GAIL-based framework was developed in \cite{9_Generative2024Peiying}, to enhance the resource allocation efficiency for VNE in dynamic SAGINs. Through adversarial training and imitating expert embedding behaviors, the GAIL generates efficient computing and bandwidth resource allocation strategies without relying on manually defined reward functions. This approach contributes to agentic learning by imitation and autonomous policy generalization, two key cognitive traits of agentic AI for operating in open-ended and cross-domain SLAETN environments. Compared to the traditional DRL-VNE, the GAIL-VNE doubles the convergence speed, reduces ground node utilization from $68\%$ to $55\%$, and increases space/air network utilization by $12\%-13\%$, significantly improving cross-domain resource allocation.

\subsubsection{Network Design and Orchestration}
Beyond dynamic decision-making and resource allocation, generative techniques, especially those incorporating LLM-based reasoning, have been increasingly adopted to tackle system-level network design and orchestration challenges in SLAETNs \cite{c14_Large2024Qiu,c15_Large2025Maatouk}. Satellite networks always suffer from inefficient and error-prone manual modeling due to their complex operational scenarios \cite{c16_Interactive2024Zhang}. To overcome this limitation, a GAI agent framework alongside an MoE-PPO method was employed in \cite{a1_Generative2024Zhang}. In this framework, the GAI agent employs LLMs with retrieval-augmented generation (RAG) to automatically model scenarios, protocols, and optimization objectives. This process reflects the reasoning and knowledge abstraction capabilities of agentic AI, enabling autonomous orchestration workflows. The simulations show that the proposed scheme achieves over $80\%$ modeling accuracy. 

\begin{figure*}[ht]
	\begin{center}
		\includegraphics[width=0.8\textwidth]{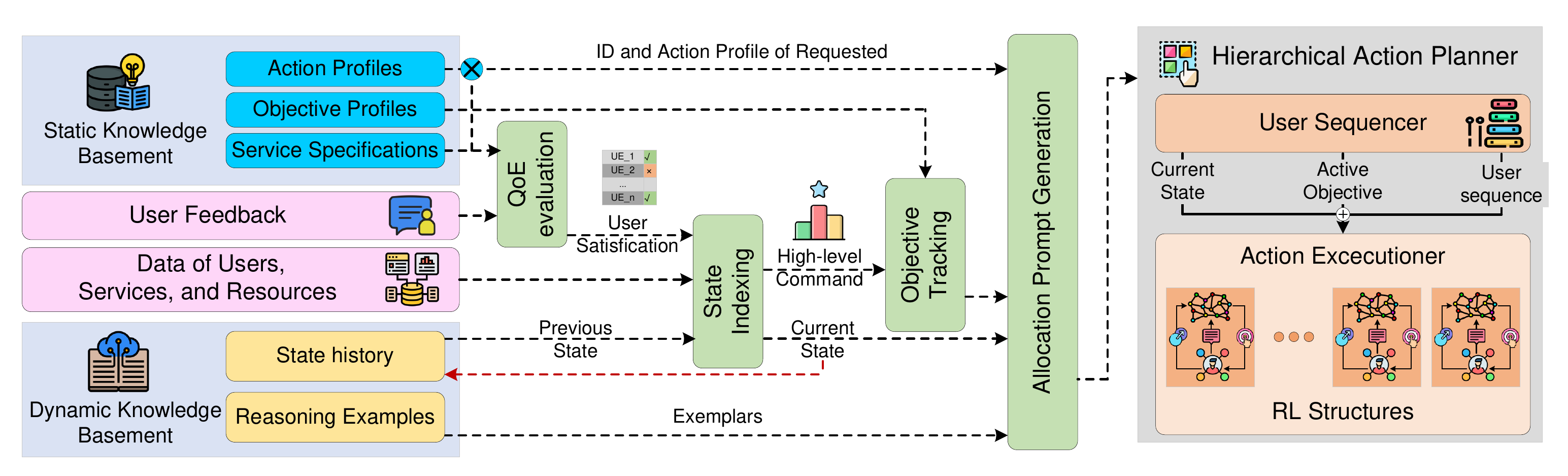}
		\caption{This illustration presents the resource allocation workflow of the ARC framework in \cite{10_An2025Masoud}, where the RAG collects system states and user feedback to generate allocation prompts. Hierarchical Action Planner then uses these prompts to derive an optimal user sequence via an LLM-based User Sequencer, and RL agents execute low-level resource allocation actions in the specified order.}
		\label{III-B-3}
	\end{center}
\end{figure*}

Based on \cite{a1_Generative2024Zhang}, the authors in \cite{10_An2025Masoud} introduced an autonomous RL control (ARC) framework to address the orchestration challenges in semantic-aware SAGINs, which is illustrated in Fig.~\ref{III-B-3}. This framework combines LLM-based CoT reasoning for high-level planning with RL agents for low-level execution, thus mitigating LLM hallucinations \cite{c52_Mitigation2024de}, reducing computational cost, and enhancing dynamic adaptability. This architectural separation demonstrates hierarchical cognition and modular decision synthesis, two hallmark traits of agentic AI systems. The simulations show that in a $10$-node semantic SAGIN scenario, this hierarchical design achieves $95\%$ cost-optimality and exhibits faster convergence and better robustness than those non-RL and reward-free baselines.

Additionally, the authors in \cite{11_Space-air-ground2025Gao} proposed a task-driven multi-agent intelligent networking (MAIN) architecture, where LLMs empower semantic-aware sensing, decision-making, and control across SAGINs. Specifically, LLMs convert high-level mission intents into executable configurations using semantic signaling and content-centric routing, embodying intent-aware orchestration and adaptive control in agentic AI. In a satellite-UAV-ship network for ocean monitoring, the MAIN-empowered duct-aware scheme reduces UAV swarm energy consumption for data feedback by $50\%$ compared to the non-duct-aware scheme, with a much gentler energy growth as transmission distance increases.

\textbf{Lesson Learned.} 
As summarized in Table~\ref{GAI in Network Management and Optimization}, the integration of GAI and LLMs significantly enhances network management and optimization within SLAETNs.  
However, many existing GAI and LLM-based methods rely heavily on domain-specific or expert-labeled data, limiting generalization across heterogeneous network conditions \cite{b18_Over-the-Air2025Fu}. Future research should focus on building lightweight and data-efficient generative frameworks that incorporate domain priors and online learning to ensure scalable and robust deployment.

\begin{table*}[ht]
\renewcommand\arraystretch{1.2}
\belowrulesep=0pt
\aboverulesep=0pt
\centering
\caption{GAI in Network Management and Optimization}
\label{GAI in Network Management and Optimization}
\begin{tabular}{>{\centering\arraybackslash}m{0.11\textwidth} | | >{\centering\arraybackslash} m{0.03\textwidth} | >{\centering\arraybackslash}m{0.08\textwidth} | >{\raggedright\arraybackslash} m{0.21\textwidth} | >{\raggedright\arraybackslash}m{0.21\textwidth}| >{\raggedright\arraybackslash} m{0.21\textwidth}}
\toprule
\midrule
    Type  & \multicolumn{1}{c|}{Ref} & Algorithm & Description & Pros & Cons \\ \toprule\midrule
\multirow{2}{*}{
  \centering
  \begin{minipage}[c][1.9cm][c]{\linewidth}
    Network Decision-Making
  \end{minipage}
}      
    & \cite{3_Routing2025Guo} & GAN-DQN & GAN-DQN is employed for efficient routing decision-making in SAGIN communications. 
    & \textbullet \ Reduced overhead of real-time data collection \newline  \textbullet~Enhanced adaptability to dynamic topologies 
    & \textbullet~Dependence on historical data quality \newline \textbullet~High computational cost of hybrid models \\ \cline{2-6} 
    & \cite{2_Generative2025Tao} & DMAPPO & DMAPPO is employed to address the multi-objective vertical handover decision problem in SAGINs. 
    & \textbullet \ Superior multi-objective optimization capability \newline  \textbullet~Robust generative capability \newline \textbullet~Improved learning efficiency and stability 
    & \textbullet~Dependence on high-quality trajectory data \newline \textbullet~High computational overhead for training \\ \hline
\multirow{3}{*}{
  \centering
  \begin{minipage}[c][3.6cm][c]{\linewidth}
    Resource Allocation
  \end{minipage}
}    
    & \cite{a3_Carrier2025Khoramnejad}& DDPM-DRL & DDPM-DRL is employed to optimize the JCALB in LEO satellites-based NTNs. 
    & \textbullet \ Superior handling of high-dimensional action spaces \newline \textbullet~Robustness to dynamic network conditions \newline \textbullet~Superior training stability and sample Efficiency
    & \textbullet~Dependence on accurate CSI \newline \textbullet~High computational overhead for training\\ \cline{2-6}
    & \cite{5_Diffusion-Enabled2025Li}& DRA & DRA is employed to support DT synchronized resource allocation in SAGINs. 
    & \textbullet \ Superior integer decision generation \newline \textbullet~Enhanced resource utilization 
    & \textbullet~Dependence on specific modeling assumptions \newline \textbullet~High computational overhead for training\\ \cline{2-6} 
    & \cite{9_Generative2024Peiying} & GAIL-VNE & GAIL-based framework is developed to enhance the resource allocation efficiency for VNE in dynamic SAGINs. 
    & \textbullet \ Reward-free learning framework \newline \textbullet~Data-efficient expert strategy imitation \newline \textbullet~Faster convergence than DRL-VNE 
    & \textbullet~Dependence on high-quality expert sample \newline \textbullet~High computational overhead for training \\ \hline
\multirow{3}{*}{
  \centering
  \begin{minipage}[c][3cm][c]{\linewidth}
    Network Design \& Orchestration
  \end{minipage}
}   
    & \cite{a1_Generative2024Zhang} & LLM & LLM-based reasoning is combined with RL for intelligent network autonomy in SAGINs. 
    & \textbullet \ Automated complex modeling \newline \textbullet~Interactive customization 
    & \textbullet~Dependence on RAG quality \newline \textbullet~High computational overhead\\ \cline{2-6} 
    & \cite{10_An2025Masoud} & LLM & LLM-based ARC is employed for addressing the orchestration challenges in semantic-aware SAGINs. 
    & \textbullet \ Improved quality of decision-making \newline \textbullet~Reduced computation overhead 
    & \textbullet~Dependence on high-quality prompts \newline \textbullet~Inadequate adaptation\\ \cline{2-6}
    & \cite{11_Space-air-ground2025Gao} & LLM & LLM-based MAIN is employed for the semantic-aware sensing, decision-making, and control across SAGINs. 
    & \textbullet \ Cross-domain orchestration \newline \textbullet~Task adaptability 
    & \textbullet~Large model size and high computational cost \newline \textbullet~Latency concerns\\ \hline
\end{tabular}
\end{table*}

\section{GAI and LLMs for SLAETN Security and Privacy Protection Enhancement}\label{sec:security}
Network security and privacy protection hinge on three core attributes including confidentiality, integrity, and availability, which are vital for trustworthy information exchange in integrated networks. However, the open architecture, cross-domain connectivity, and dynamic topology of SLAETNs significantly amplify secure risks, such as eavesdropping, spoofing, intrusion, and service disruption. This section explores how the integration of GAI and LLMs enables agentic AI's capabilities in confident and covert communications, intelligent threat detection, and adaptive defense and recovery, thereby enhancing the overall resilience of SLAETN security and privacy.

\subsection{Communication Confidentiality and Covertness}

Communication confidentiality and covertness are essential for protecting network security and user privacy, especially in dynamic environments like SLAETNs. Confidentiality ensures that transmitted data remains private, while covertness hides the presence of communication \cite{b16_Multi-Objective2025Zhang,b33_Generative2024Liao}. Despite traditional cryptographic methods, PLS, and covert communication techniques show promise in enhancing the network security \cite{r8_Joint2025Zhang}, they mainly depend on static assumptions or handcrafted features \cite{19_Privacy-Preserving2025Li,r13_Secrecy2025Gao}. These limitations reduce their effectiveness in complex, real-world conditions and hinder adaptability to adversarial attacks \cite{20_Hash-Chain-Based2020Luo}. Currently, GAI brings new opportunities through its ability to learn representations, model uncertainty, and optimize decisions dynamically. In particular, by leveraging deep generative models and neural policies, GAI methods, such as GANs, TBMs, and VAEs, have demonstrated remarkable potential in improving secrecy energy efficiency (SEE) optimization \cite{4_Toward2025Kakati, 24_Low-altitude2025Jia}, enhancing authentication \cite{a25_Radio2024Jiang}, and supporting robust covert communications \cite{23_Covert2024Jia,a18_Res-GAN2024Wang}, which offers a promising path toward intelligent and secure wireless communication.

\subsubsection{Secrecy Energy Efficiency Optimization}
The prevalent LoS channels between LEO satellites and UAVs in SLAETNs are highly vulnerable to eavesdropping, inducing critical information leakage risks \cite{c20_Joint2022Han}. Meanwhile, energy-constrained aerial platforms necessitate joint trajectory-power optimization for sustainable operation, yet existing methods predominantly optimize secrecy rate (SR) \cite{c22_Broadcast2023Han,r3_Robust2025Zhang} or energy efficiency \cite{c21_Deep2024Wu,r6_Dynamic2025Li} in isolation, neglecting the crucial secrecy-energy interplay under dynamic space-air-ground environments. To overcome this limitation, the authors in \cite{4_Toward2025Kakati} developed a GAN-empowered DRL framework for non-convex SEE optimization. In this framework, the GAN generator generates action policies, such as HAP trajectories, user associations, beamforming, and channel predictions. While the discriminator ensures the realism and feasibility of policies under network constraints. This integration of GAN with DRL demonstrates agentic capabilities in policy synthesis and feasibility-aware decision modeling, enabling precise prediction and rapid adaptation to SLAETNs’ spatiotemporal dynamics. The simulation shows that GAN-DRL enhances SEE by $7.85\%$ and $12.43\%$ compared to long short-term memory (LSTM)-DDQN and DDQN, respectively, achieving only $5.1\%$ gap from the exhaustive  search solution. 

Additionally, the authors in \cite{24_Low-altitude2025Jia} proposed a transformer-enhanced soft-actor-critic (TransSAC) algorithm for dynamic satellite-vessel communication scenarios to balance SR and UAV energy consumption. The TransSAC employs a self-attention mechanism to capture cross-temporal dependencies in state-action interactions and to generate adaptive friendly-jamming power control and UAV 3D trajectory strategies. This combination of TBMs and SAC overcomes the limitation of traditional RL that relies only on the current state \cite{c53_Generative2025Sun}, enhancing long-horizon reasoning and strategic planning which are key components of agentic AI in energy-aware secure networking. Through simulations, compared to the conventional RL algorithms such as deep deterministic policy gradient (DDPG), twin delayed DDPG (TD3), PPO, and SAC, TransSAC improves the SR by $10\%-23\%$ and reduces energy consumption by $14\%-33\%$, mitigating the local optimum problem of traditional DRL in long-time optimization. The works in \cite{4_Toward2025Kakati} and \cite{24_Low-altitude2025Jia} primarily address communication confidentiality by enhancing secure transmission.

\subsubsection{Radio Frequency Fingerprint Identification}
Moreover, radio frequency fingerprint identification (RFFI) is also an effective PLS solution for communication confidentiality, which achieves physical layer authentication through transmitter hardware defects \cite{c18_A2020Soltanieh}. However, most existing RFFI methods predominantly rely on manual features and are constrained by prior assumptions of linear transformations, thus struggling to fully extract the nonlinear characteristics of RF fingerprints \cite{c19_A2021Wang}. In contrast, VAE models achieve data compression through maximization of the variational lower bound while preserving feature diversity, thereby overcoming the inherent limitations of linear dimensionality reduction methods such as principal component analysis (PCA) \cite{c54_Efficient2025Pandit}.
For example, in response to interference threats faced by the global navigation satellite system (GNSS), the authors in \cite{a25_Radio2024Jiang} proposed a VAE-LSTM based radio frequency fingerprinting scheme for PLS authentication. In this scheme, with LSTM networks acting as the encoder and decoder to extract temporal features from post-correlation GNSS signals, a VAE is employed to generate denoised and nonlinearly compressed representations by learning and sampling from their latent distributions. This mechanism enables agentic feature abstraction and nonlinear perception, allowing the system to robustly distinguish identities under noisy and uncertain signal environments. Numerical results on Nuremberg GPS datasets show that this VAE-LSTM scheme achieves an identification accuracy of $95.68\%$, while reducing computational complexity by $58\%$ compared to the conventional PCA-LSTM method. This lightweight solution strengthens communication confidentiality through effective identity verification.

\subsubsection{Covert Communications}
While confidentiality protects the content of communication, covertness further ensures that the existence of communication remains undetectable, thus serving as a higher-layer enhancement to traditional confidentiality \cite{c17_Secrecy2020Wang}. 
For the covert communications in LEO satellite systems with imperfect CSI, the authors in \cite{23_Covert2024Jia} proposed a data-driven GAN (DD-GAN)-based cooperative UAV jamming scheme. In the proposed scheme, the generator of GAN produces UAV jamming power allocation and trajectory scheduling strategies, while the discriminator simulates eavesdropper detection behavior. This adversarial interaction supports agentic policy generation under partial observability, enabling robust optimization without full prior knowledge. Through simulations, the DD-GAN achieves an effective trade-off between covert rate and detection error probability with limited prior information, outperforming traditional genetic algorithms requiring full eavesdropper information. 

Furthermore, the authors in \cite{a18_Res-GAN2024Wang} proposed a ResGAN model for behavioral modeling of satellite power amplifiers (PAs). This model effectively suppresses out-of-band spectral leakage, demonstrating the potential of GAI to enhance communication covertness by reducing signal detectability. Specifically, by leveraging adversarial training to extract nonlinear amplifier characteristics, the proposed Res-GAN contributes to agent-level behavioral reasoning and signal shaping. Residual connections are incorporated into the generator to address gradient vanishing and accelerate convergence, thereby improving model accuracy and stability. Compared to conventional LSTM and deep neural network (DNN) schemes, the proposed Res-GAN scheme improves normalized MSE by approximately $2\,\text{dB}$ and reduces adjacent channel power ratio by $2-4\,\text{dBc}$. Through improved PA linearization and reduced interference leakage, this work enhances signal integrity and covertness.

\textbf{Lesson Learned.} 
In conclusion, GAI models have exhibited excellent performance in communication confidentiality and covertness enhancement within SLAETNs, as summarized in Table~\ref{GAI in Communication Confidentiality and Covertness}. However, challenges remain in model complexity, data dependency, and environmental adaptability. Future work should focus on lightweight, robust, and adaptive designs of for real-time secure transmission.

\begin{table*}[ht]
\renewcommand\arraystretch{1.2}
\belowrulesep=0pt
\aboverulesep=0pt
\centering
\caption{GAI in Communication Confidentiality and Covertness}
\label{GAI in Communication Confidentiality and Covertness}
\begin{tabular}{>{\centering\arraybackslash}m{0.11\textwidth} | | >{\centering\arraybackslash} m{0.03\textwidth} | >{\centering\arraybackslash}m{0.08\textwidth} | >{\raggedright\arraybackslash} m{0.21\textwidth} | >{\raggedright\arraybackslash}m{0.21\textwidth}| >{\raggedright\arraybackslash} m{0.21\textwidth}}
\toprule
\midrule
    Type  & \multicolumn{1}{c|}{Ref} & Algorithm & Description & Pros & Cons \\ \toprule\midrule
\multirow{2}{*}{
  \centering
  \begin{minipage}[c][2.2cm][c]{\linewidth}
    SEE Optimization
  \end{minipage}
}      
    &\cite{4_Toward2025Kakati} & GAN-DRL & GAN-DRL is employed for non-convex SEE optimization in space-air-ground communications. 
    & \textbullet \ Dynamic adaptation to non-convexity \newline  \textbullet~Enhanced temporal dependency modeling 
    & \textbullet~High computational overhead for training \newline \textbullet~Vulnerability to high-dimensional environments \newline \textbullet~Reliance on accurate CSI\\ \cline{2-6} 
    &\cite{24_Low-altitude2025Jia} & TransSAC & TransSAC is employed to balance SR and UAV energy consumption for dynamic satellite-vessel communication scenarios. 
    & \textbullet \ Superior handling of long-term dependencies \newline  \textbullet~Dynamic multi-objective weight exploration \newline \textbullet~Computational efficiency via offline learning 
    & \textbullet~High computational overhead for training \newline \textbullet~Reliance on perfect CSI\\ \hline
\multirow{1}{*}{
  \centering
  \begin{minipage}[c][0.45cm][c]{\linewidth}
    RFFI
  \end{minipage}
}    
    & \cite{a25_Radio2024Jiang} & VAE-LSTM & VAE-LSTM is employed for lightweight RFFI in GNSS systems. 
    & \textbullet \ Enhanced noise robustness \newline \textbullet~Superior feature representation \newline \textbullet~Improved computational efficiency
    & \textbullet~Dependence on signal phase consistency \newline \textbullet~Limited Spoofer Diversity \\  \hline
\multirow{2}{*}{
  \centering
  \begin{minipage}[c][1.7cm][c]{\linewidth}
    Covert Communications
  \end{minipage}
}   
    & \cite{23_Covert2024Jia} & DD-GAN & DD-GAN is employed for jointly UAV trajectory and jamming power optimization in covert communication scenarios. 
    & \textbullet \ Robustness under partial information \newline \textbullet~Data-driven adaptability 
    &\textbullet~Dependence on training data quality \newline \textbullet~High computational overhead for training\\ \cline{2-6} 
    & \cite{a18_Res-GAN2024Wang} & Res-GAN & Res-GAN is employed for behavioral modeling of satellite PAs. 
    & \textbullet \ Enhanced modeling accuracy \newline \textbullet~Improved training stability 
    & \textbullet~Sensitivity to training data quality and quantity \newline \textbullet~Limited adaptation to rapid PA characteristic changes
\\ \hline
\end{tabular}
\end{table*}

\subsection{Anti-Spoofing and Intrusion Detection}

Anti-spoofing and intrusion detection are essential to ensuring entity authenticity and access integrity, forming the first line of defense in safeguarding network security and privacy in SLAETNs. Anti-spoofing targets identity forgery by detecting illegitimate entities posing as trusted nodes, whereas intrusion detection focuses on identifying abnormal behaviors indicative of unauthorized access. Unlike confidentiality- and covertness- focused mechanisms that emphasize protecting the content of transmission, these two techniques aim to protect network integrity. The former emphasizes identity validation \cite{a27_Variational2024Zhang}, while the latter concentrates on behavioral and system anomaly analysis \cite{a12_Intrusion2024Jiang}. However, conventional methods, including statistical thresholds, rule-based engines, and feature-engineered classifiers, often fail to adapt to stealthy, evolving, and cross-domain attacks, especially under the dynamic and heterogeneous conditions of SLAETNs. To address these challenges, GAI and LLMs offer a promising alternative by leveraging deep generative priors, semantic reasoning, and threat-informed data augmentation. This enables more context-aware, adaptive, and scalable anti-spoofing and intrusion detection tailored to the unique security demands of agentic SLAETNs.

\subsubsection{Anti-Spoofing}
Satellite systems, particularly in global navigation, have seen widespread adoption. However, the open-access nature and low transmission power of civilian signals expose them to spoofing attacks \cite{c29_Decentralized2018Milaat}. These attacks typically transmit high-power signals structurally similar to authentic ones \cite{c30_GPS2012Jahromi}, misleading receivers and causing incorrect timing or positional outputs. Traditional detection techniques often rely on identifying multiple correlation peaks in the acquisition matrix, yet they fail when spoofing signals are tightly synchronized with authentic ones, resulting in indistinguishable single-peak profiles \cite{c31_Low-complexity2018Tu}. Addressing this issue, a GNSS anti-spoofing scheme based on the GAN was introduced in \cite{a32_GNSS2021Li}, to achieve effective spoofing detection even when the deceptive signals are highly synchronized with authentic ones, thereby strengthening communication integrity. Specifically, the GAN generator synthesizes signal-like data during training to help the discriminator learn the statistical distribution of genuine GNSS correlation matrices. This adversarial learning process supports perceptual generalization of agentic AI, equipping the model to recognize spoofing even when traditional correlation-based features are suppressed. Through simulations, the effectiveness of the proposed scheme is notable, with a spoofing detection rate exceeding $98\%$ when the code phase delay is over $0.5\%$ chip, significantly outperforming the conventional CNN-based scheme.

Beyond synchronized spoofing, the growing prevalence of zero-day attacks that are enabled by open-source exploit tools introduces further challenges. Conventional supervised learning methods, which rely on predefined attack signatures, often fail to detect such unknown threats, especially under strict latency constraints \cite{c32_GNSS2018Wesson}. Motivated by \cite{c33_Unsupervised2021Zoppi}, the authors in \cite{a24_Machine2024Iqbal} proposed a VAE-based zero-day attacks detection method to reinforce integrity under unseen threats. By generating reconstructed GNSS signal features from a learned latent distribution, VAE models the normal behavior of authentic signals to enable spoofing detection via reconstruction error. This unsupervised strategy embodies latent-space reasoning and anomaly awareness of agentic AI, enabling the model to detect unseen attack patterns without prior labeling. When evaluated on the DS-7 complex attack scenario from the TEXBAT dataset, the VAE-based scheme achieved a detection probability of over $92.5\%$. This represents a $41.8\%$ improvement compared to the supervised learning-based two-stage artificial neural network scheme.

\begin{figure}[!t]
	\begin{center}
		\includegraphics[width=1\columnwidth]{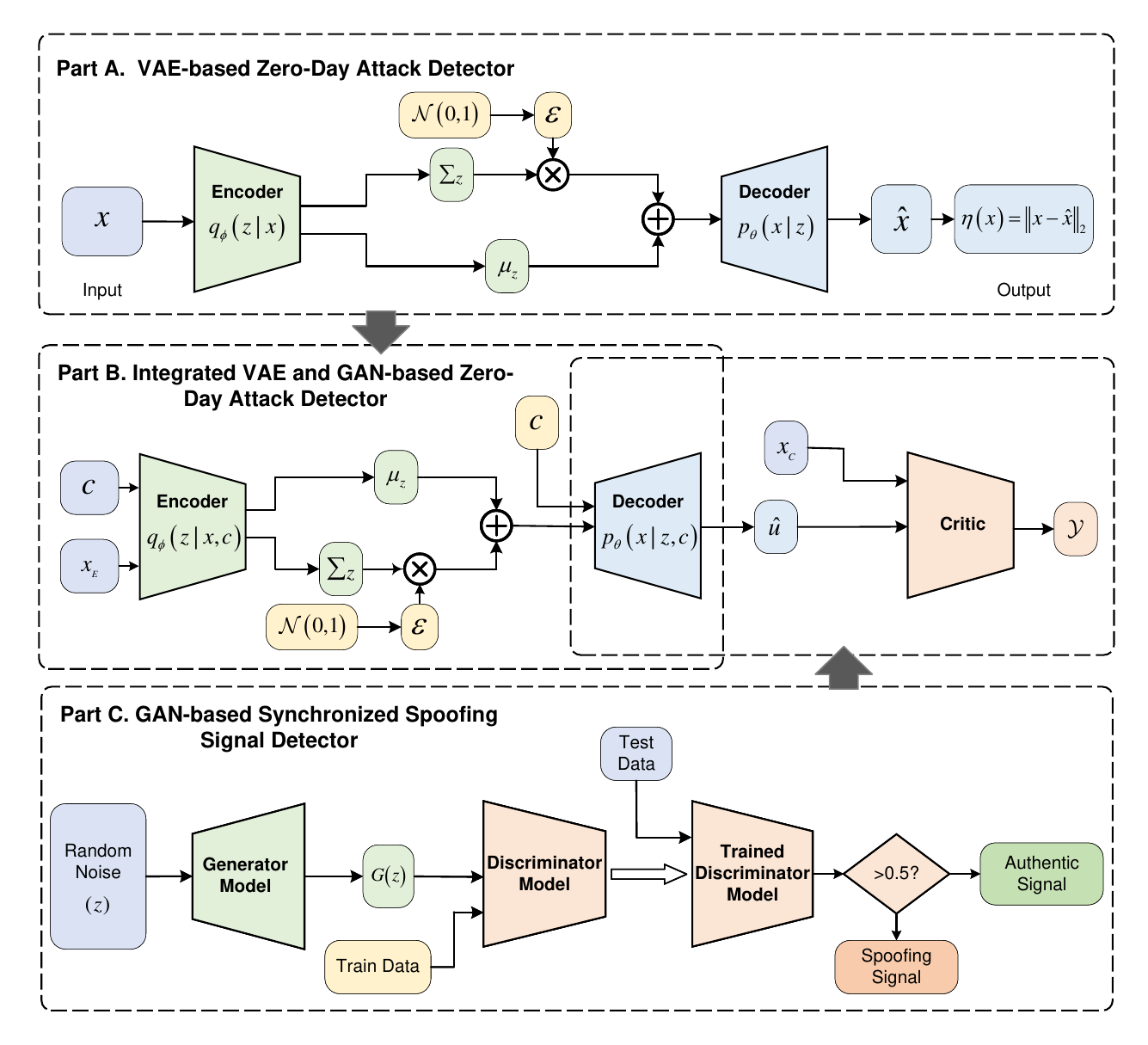}
		\caption{This illustration presents the three spoof detection architectures of \cite{a32_GNSS2021Li,a24_Machine2024Iqbal,a26_A2024Iqbal}. Part A illustrates the VAE-based architecture in \cite{a24_Machine2024Iqbal}. Part C illustrates the GAN-based architecture in \cite{a32_GNSS2021Li}, which uses the trained discriminator to distinguish spoofing signal. In part B, the integrated model in \cite{a26_A2024Iqbal} utilizes both VAE and GAN for more effective and stable attack detection.}
		\label{IV-B-1}
	\end{center}
\end{figure}

To leverage the complementary strengths of VAEs and GANs, the authors further proposed a unified VAE-GAN architecture for detecting GNSS spoofing attacks in \cite{a26_A2024Iqbal}, with particular emphasis on zero-day threats. The VAE-GAN first generates reconstructed GNSS signals from a latent representation using the VAE decoder. These reconstructions are then evaluated by a GAN critic to model the distribution of authentic signals, enabling spoofing detection based on reconstruction fidelity and critic feedback. This combination enhances self-consistency validation of agentic AI, by aligning reconstruction quality with learned authenticity criteria. Additionally, the hybrid framework mitigates the blurriness of VAE outputs and stabilizes GAN training. Notably, compared to the traditional supervised learning-based scheme, the proposed VAE-GAN improves the detection probability by over $50\%$ under the DS-7 complex attack scenarios, effectively enhancing identity integrity. Fig.~\ref{IV-B-1} presents the GAI model architectures of the aforementioned three studies \cite{a32_GNSS2021Li,a24_Machine2024Iqbal,a26_A2024Iqbal} that are focused on anti-spoofing.

\subsubsection{Intrusion Detection}
Due to the open nature of wireless channels and the dynamic network topology, STINs are more vulnerable to intrusions than conventional networks, leading to severe economic and security risks. However, centralized detection approaches are inadequate for processing large-scale, heterogeneous data from satellite and terrestrial sources, and they also pose privacy concerns \cite{c34_Distributed2020Li}. As a remedy, FL enables distributed and privacy-preserving intrusion detection at local nodes \cite{c35_Attack2020Wang}. Nevertheless, its effectiveness diminishes under non-independent and identically distributed (non-IID) traffic conditions, leading to reduced accuracy in detecting both satellite-specific and terrestrial attacks \cite{c36_Federated2021Zhu}.
To address this issue, the authors in \cite{a48_A2025He} proposed a multi-tier FL framework with a differentially private CDM (DP-CDM). This framework promotes integrity by enabling robust anomaly detection and also enhances confidentiality through differential privacy preservation. Specifically, the DP-CDM learns local traffic distributions through a forward-noising and reverse-denoising process, to generate synthetic data that approximates the global distribution within the STIN. This process enables agentic distribution-aware perception and uncertainty-adaptive generation, mitigating blind spots introduced by non-IID data. During training, dynamic feature selection and differential privacy mechanisms protect sensitive data. Each node then combines global synthetic and local real data to train its intrusion detection model. The experiments show that the DP-CDM scheme achieves $96.63\%$ accuracy under non-IID settings, with a $2.41\%$ gain over the state-of-the-art baseline.

In addition, satellite networks face stealthy cyber threats that mimic legitimate traffic to evade signature-based intrusion detection systems. These attacks often employ adaptive obfuscation and distributed evasion strategies that conventional ML models may fail to capture \cite{c37_ICARUS2021Giuliari}. To enhance STIN cyber security, the authors in \cite{a56_PLLM-CS2024Mohammed} introduced a transformer-based pre-trained LLM (PLLM) architecture. By generating contextual token representations from multivariate satellite network data, the PLLM enables the transformer to capture long-range semantic dependencies, thereby supporting agentic reasoning and context-aware classification under adversarial conditions. Impressively, the proposed PLLM achieves $100\%$ accuracy on the UNSW-NB15 dataset and outperforms CNN and BiLSTM on the TON-IoT benchmark, demonstrating its effectiveness in detecting network intrusion and improving traffic integrity.

\textbf{Lesson Learned.} 
The incorporation of GAI and LLMs into anti-spoofing and intrusion detection tasks in Table~\ref{GAI in Anti-Spoofing and Intrusion Detection} demonstrates substantial potential in addressing the limitations of traditional detection paradigms in SLAETNs. By generating authentic-like signals and learning semantic anomalies, GAI and LLMs empower anti-spoofing and intrusion detection, thereby improving SLAETN integrity. However, current methods still suffer from high training cost, data dependency, and limited adaptability to dynamic cross-domain environments. Future work should focus on developing lightweight and robust models that generalize across scenarios and support real-time defense.

\begin{table*}[ht]
\renewcommand\arraystretch{1.2}
\belowrulesep=0pt
\aboverulesep=0pt
\centering
\caption{GAI in Anti-Spoofing and Intrusion Detection}
\label{GAI in Anti-Spoofing and Intrusion Detection}
\begin{tabular}{>{\centering\arraybackslash}m{0.11\textwidth} | | >{\centering\arraybackslash} m{0.03\textwidth} | >{\centering\arraybackslash}m{0.08\textwidth} | >{\raggedright\arraybackslash} m{0.21\textwidth} | >{\raggedright\arraybackslash}m{0.21\textwidth}| >{\raggedright\arraybackslash} m{0.21\textwidth}}
\toprule
\midrule
    Type  & \multicolumn{1}{c|}{Ref} & Algorithm & Description & Pros & Cons \\ \toprule\midrule
\multirow{3}{*}{
  \centering
  \begin{minipage}[c][3.5cm][c]{\linewidth}
    Anti-Spoofing Detection
  \end{minipage}
}  
    & \cite{a32_GNSS2021Li} & GAN & GAN is employed for effective GNSS spoofing jamming detection. 
    & \textbullet \ Effective feature learning \newline  \textbullet~Generalization across scenarios
    & \textbullet~Dependence on sufficient training data \newline \textbullet~High computational overhead for training \newline \textbullet~Lack of real-world validation \\ \cline{2-6} 
    & \cite{a24_Machine2024Iqbal} & VAE & VAE-based zero-day attack detection method is employed for improving the resilience to unknown GNSS spoofing. 
    & \textbullet \ Independence on attack signatures \newline  \textbullet~Data-driven feature representation
    & \textbullet~Dependence on sufficient genuine data \newline \textbullet~High computational overhead for training \\ \cline{2-6} 
    & \cite{a26_A2024Iqbal} & VAE-GAN & VAE-GAN is employed for detecting GNSS spoofing attacks, particularly zero-day attacks. 
    & \textbullet \ Effective representation learning with limited data \newline  \textbullet~Robustness against subtle attacks \newline \textbullet~Generalization to diverse attack scenarios
    & \textbullet~Dependence on high-quality genuine training data \newline \textbullet~High computational overhead for training \newline \textbullet~Training instability and hyperparameter sensitivity \\ \hline

\multirow{2}{*}{
  \centering
  \begin{minipage}[c][2.3cm][c]{\linewidth}
    Intrusion Detection
  \end{minipage}
}
    & \cite{a48_A2025He} & DP-CDM & DP-CDM is employed to to generate high-quality synthetic intrusion samples for balancing the privacy protection and detection performance in STINs. 
    & \textbullet \ Effective handling of Non-IID data \newline \textbullet~Stability and high-quality generation \newline \textbullet~FL integration 
    & \textbullet~High computational overhead for training \newline \textbullet~Sensitivity to dynamic network topology \newline \textbullet~Dependence on labeled data for conditioning \\ \cline{2-6} 
    & \cite{a56_PLLM-CS2024Mohammed} & PLLM & PLLM is employed to enable semantic-aware and cross-scenario cyber threat detection in satellite networks. 
    & \textbullet \ Effective handling of adaptive cyber threats \newline \textbullet~Generalization across network domains \newline \textbullet~Efficient long-range dependence modeling
    & \textbullet~Lack of dedicated satellite network datasets \newline \textbullet~Computational overhead for edge deployment\\ \hline
\end{tabular}
\end{table*}

\subsection{Adaptive Security Defense and Signal Recovery}

Adaptive security defense constitutes a critical component in ensuring network availability and resilience by enabling real-time threat detection \cite{c23_TEL-DT2023Guo}, situational awareness \cite{c24_A2021Chen}, and agile countermeasures against evolving cyber-physical attacks \cite{c15_Large2025Maatouk}. Once such threats impair the communication process, signal recovery techniques become essential for reconstructing corrupted or incomplete transmissions, thereby preserving data integrity and sustaining system operability\cite{c26_Image2018Zhu}. Traditional solutions, such as rule-based intrusion detection, static anomaly models, and sparsity-driven signal reconstruction, often exhibit limited adaptability, high complexity, and degraded performance under adversarial or nonstationary conditions \cite{a36_Transformer-based2024Manuel}. In this context, generative techniques have emerged as a promising paradigm capable of modeling intricate data distributions and generating high-fidelity representations under uncertainty \cite{a58_SatGuard2025Xiao,a47_Conditional2024Zhang}. By integrating generative learning into adaptive defense and signal recovery, GAI models and LLMs support proactive threat mitigation and robust signal recovery. These capabilities enhances SLAETN security in terms of availability and integrity.

\subsubsection{Security Defense Framework}
LLMs have demonstrated substantial potential in enhancing security defense frameworks for SLAETNs, especially under the zero-trust paradigm and dynamic threat environments. In \cite{8_Exploring2025Cao}, the authors proposed a foundational LLM-based situation awareness (LLM-SA) architecture that leverages LLMs for real-time threat assessment and defense policy generation. By leveraging the chain-of-thought (CoT) reasoning capability of LLMs, the proposed architecture enables dynamic threat evaluation and policy generation. This showcases the agentic capability for contextual reasoning and adaptive decision-making in adversarial settings.  Experimental results show that LLM-SA outperforms traditional autoencoder-DNN (AEDNN) \cite{c59_A2021Yang} and network attack behavior classification (NABC) \cite{c60_Network2022Yang} based approaches in terms of MSE, RMSE, and accuracy. Specifically, the LLM-SA scheme based on Llama3-8B achieves optimal defense performance for zero-trust SAGINs, with an MSE of $0.07$ and an accuracy of $91.7\%$. Notably, it achieves this while using only $0.44\%$ of ChatGPT-4’s parameters, demonstrating high adaptability and lightweight deployment, which is key to maintaining network availability in large-scale SAGIN systems, like SLAETNs.

To further adapt to unknown threats, the authors in \cite{31_An2025Qi} presented a self-evolving security framework for 6G SAGINs. This framework integrates an LLM-based ``6G-network guard'' for threat interpretation and defense strategy generation, alongside a ``6G-instruction'' module for automatic knowledge update and continual learning. These features embody agentic self-refinement and evolving policy cognition, critical for zero-trust environments. Built upon Llama3-8B, the system supports lightweight deployment and achieves real-time response with a latency of just $127\,\text{ms}$. Experimental results on three datasets demonstrate that the proposed framework improves policy accuracy by an average of $50.15\%$ compared to state-of-the-art LLMs, such as GPT-3.5 and GPT-4. Besides, this framework maintains robust detection performance under SQL injection and DDoS scenarios, ensuring system availability even under unknown attacks.
Collectively, these studies in \cite{8_Exploring2025Cao} and \cite{31_An2025Qi} demonstrate how LLM-based models, when equipped with structured reasoning and continual adaptation mechanisms, enable intelligent, scalable, and agentic security orchestration for SLAETNs.

\subsubsection{Signal Recovery}
Reliable signal recovery plays a pivotal role in safeguarding SLAETN security and privacy, as it ensures the integrity and availability of legitimate communication signals in the presence of interference, eavesdropping, and intentional jamming. 
When signals are corrupted or lost due to adversarial actions or harsh environments, effective recovery mechanisms help restore communication continuity and support accurate threat detection at higher protocol layers.
Despite traditional signal recovery methods, such as tensor decomposition \cite{c27_Visual2016Liu} and model-based inference \cite{c28_Learning2021Tang}, have been widely used, they often suffer from limited adaptability and degraded performance under nonstationary noise and incomplete observations. In contrast, GAI-based models exhibit significant potential in addressing signal recovery challenges in SLAETNs, particularly under constrained spectral resources and uncertain environments. 

By learning latent probabilistic representations, VAEs enable efficient compression and robust reconstruction of heterogeneous signals, even in the presence of noise or missing information. Based on this, a VAE-empowered DL framework was adopted in \cite{21_Efficient2023Jing} to enable efficient generation of fused communication and sensing data. Specifically, CNN-based encoders are used to extract feature representations from communication signals and synthetic aperture radar (SAR) images, which are then fused and mapped into a compact latent vector by the VAE. This latent vector, transmitted with low overhead, is decoded to reconstruct the original data. This approach reflects agentic perceptual abstraction and resilient signal synthesis, allowing accurate and efficient recovery under bandwidth and noise constraints.
Through simulations, the proposed method yields satisfactory recovery of communication signals and high-fidelity reconstruction of SAR images, with the MSEs of communication signals and SAR images at a SNR of $20\,\text{dB}$ as low as $0.01442$ and $0.00472$, respectively. This validates the effectiveness of VAE for communication integrity and availability protection in resource-constrained environments. 

Besides, by leveraging the strong capabilities in learning the complex distributions of time–frequency signals, GANs are also well-suited for signal recovery tasks in SLAETNs. For instance, in \cite{22_High-precision2023Guo}, a GAI-empowered reconstruction method, termed multi-component time (MTS)-GAN, was proposed for high-precision electromagnetic environment data reconstruction in space-air-ground integrated IoTs. In this MTS-GAN, a gate recursive element is introduced to help the GAN extract joint time–frequency features and simulate temporal anomalies, enabling agentic temporal reasoning and structure-aware completion under missing-data scenarios. The experiment results show that with $80\%$ missing data, the RMSE of MTS-GAN only increases by $0.1014$, which is much lower than the $0.9196$ of conventional fast low-rank tensor completion (FaLRTC) scheme \cite{c61_Tensorf2022Chen}, indicating robust communication integrity under severe data loss and high utility in availability-critical recovery tasks.

In parallel, GDMs have recently emerged as powerful tools for signal recovery in SLAETNs, offering robustness against complex interference and high-dimensional noise. For example, a diffusion model-based signal recovery scheme was proposed in \cite{a43_Diffusion2024Adam} to address the interference challenges brought by satellite-terrestrial spectrum sharing. The model uses iterative denoising to progressively refine signal structure while eliminating boundary artifacts from CNN padding, thereby supporting agentic reconstruction. Compared to the conventional CNN-based scheme, the proposed scheme reduces the normalized MSE by $0.5\,\text{dB}$ under low SNR and improves symbol error rate by approximate $4.1\,\text{dB}$ under medium/high SNR, demonstrating its reliability in restoring communication integrity under shared-spectrum interference. 

However, conventional stable diffusion models rely on variational encoders, which face performance bottlenecks in high-dimensional spaces. 
To mitigate this, a Gaussian splatting-synergized stable diffusion (GS3D) model was proposed in \cite{25_GS3D2025Adam} to reconstruct signals at Earth stations in motions within 3D wireless SAGIN environments. Specifically, the GS3D model introduces Gaussian splatting to replace the traditional variational encoder, encoding noisy signals into latent representations with reduced computational complexity. These compressed representations are then denoised by stable diffusion steps, with the generation process guided by rewards from an SAC module. This model reflects agentic reward-aware reconstruction and low-complexity latent reasoning, enabling efficient signal recovery in 3D mobile SAGIN settings. The simulations show that the GS3D achieves a normalized MSE of $10^{-3}$ and BER of $10^{-4}$, which significantly outperforms conventional DDPMs by one order of magnitude, verifying its effectiveness in improving data integrity.

\begin{figure*}[ht]
	\begin{center}
		\includegraphics[width=0.9\textwidth]{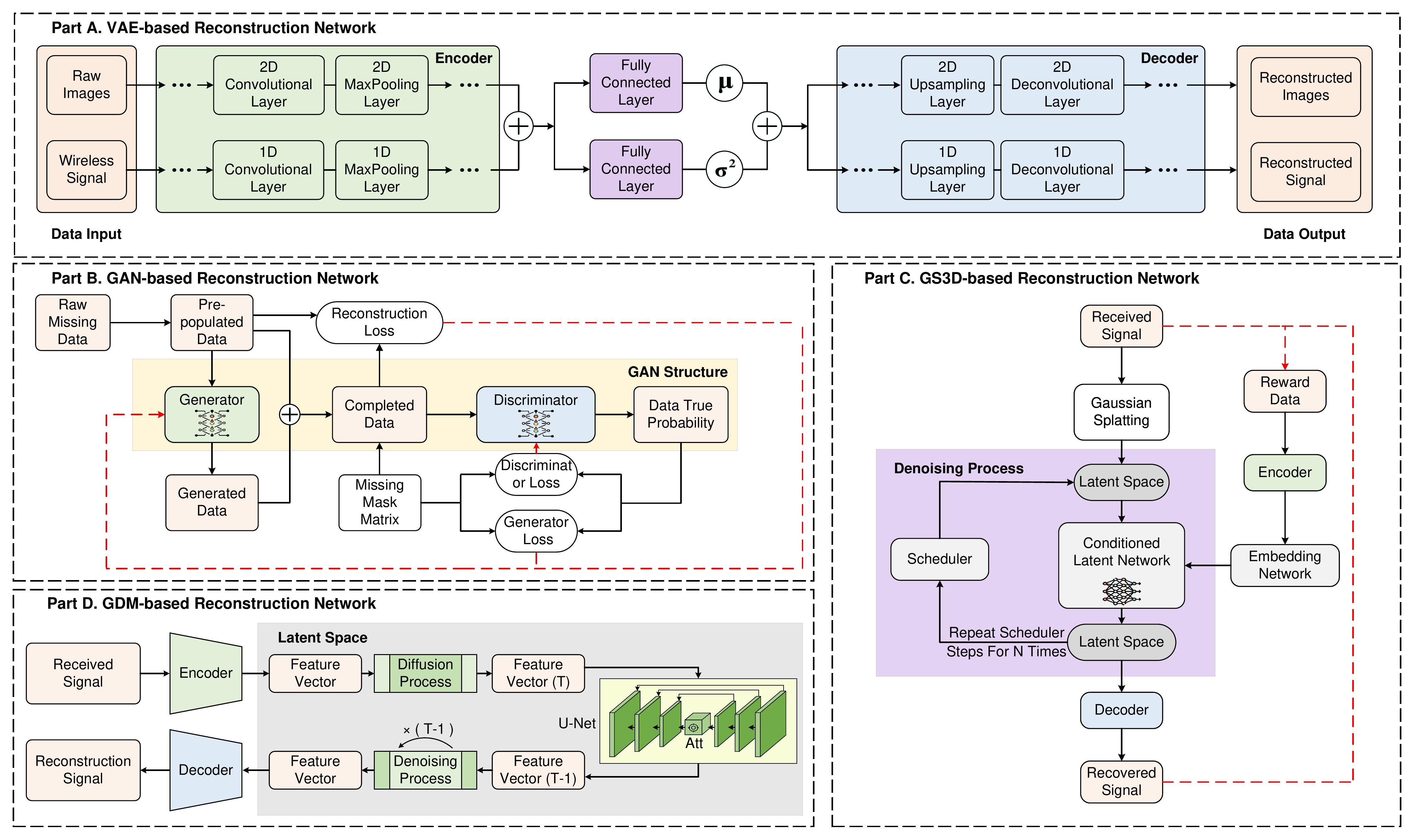}
		\caption{This illustration demonstrates the GAI-based approaches for signal reconstruction in \cite{21_Efficient2023Jing,22_High-precision2023Guo,a43_Diffusion2024Adam,a25_Radio2024Jiang}. Part A presents the VAE-based architecture in \cite{21_Efficient2023Jing} which fuses and reconstructs communication signals and SAR images with the CNN encoder and the De-CNN decoder.  Part B presents the GAN-based architecture in \cite{22_High-precision2023Guo} that generates complete electromagnetic data via adversarial training mechanism. Part C  presents the GS3D-based signal recovery model \cite{25_GS3D2025Adam} that uses Gaussian splatting and a latent UNet with reward embedding. Part D presents the GDM-based signal recovery model in \cite{a43_Diffusion2024Adam} with encoder, attention-integrated U-Net, and decoder.}
		\label{IV-C-2}
	\end{center}
\end{figure*}

\textbf{Lesson Learned.} 
As summarized in Table~\ref{GAI in Adaptive Security Defense and Signal Recovery}, GAI and LLMs enhance SLAETN's resilience by supporting adaptive threat response and high-fidelity signal reconstruction. Yet, challenges remain due to heavy computation and limited adaptability to dynamic scenarios. Future efforts should pursue efficient and transferable models to ensure timely defense and recovery.

\begin{table*}[ht]
\renewcommand\arraystretch{1.2}
\belowrulesep=0pt
\aboverulesep=0pt
\centering
\caption{GAI in Adaptive Security Defense and Signal Recovery}
\label{GAI in Adaptive Security Defense and Signal Recovery}
\begin{tabular}{>{\centering\arraybackslash}m{0.11\textwidth} | | >{\centering\arraybackslash} m{0.03\textwidth} | >{\centering\arraybackslash}m{0.08\textwidth} | >{\raggedright\arraybackslash} m{0.21\textwidth} | >{\raggedright\arraybackslash}m{0.21\textwidth}| >{\raggedright\arraybackslash} m{0.21\textwidth}}
\toprule
\midrule
    Type  & \multicolumn{1}{c|}{Ref} & Algorithm & Description & Pros & Cons \\ \toprule\midrule
\multirow{2}{*}{
  \centering
  \begin{minipage}[c][2.2cm][c]{\linewidth}
    Security Defense Framework
  \end{minipage}
}  
    & \cite{8_Exploring2025Cao} & LLM-SA & LLM-SA leverages LLMs for real-time threat assessment and defense policy generation. 
    & \textbullet \ Superior cross-domain threat processing \newline  \textbullet~Efficient lightweight architecture \newline \textbullet~Dynamic adaptive assessment
    & \textbullet~Dependence on sufficient training data \newline \textbullet~High computational overhead for complex scenarios\\ \cline{2-6} 
    & \cite{31_An2025Qi} & LLM-6GNG & LLM-6GNG is employed for threat information processing and strategy generation. 
    & \textbullet \ Dynamic strategy generation for complex threats \newline  \textbullet~Self-evolution capability via 6G-INST \newline \textbullet~Real-time efficiency with lightweight models 
    & \textbullet~Dependence on sufficient and diverse training data \newline \textbullet~Computational overhead for fine-tuning \newline \textbullet~Performance variability across attack types \\ \hline

\multirow{4}{*}{
  \centering
  \begin{minipage}[c][6cm][c]{\linewidth}
    Signal Recovery
  \end{minipage}
}
    & \cite{21_Efficient2023Jing} & VAE-DL & VAE-DL is employed to jointly fuse and reconstruct communication signals and SAR images. 
    & \textbullet \ Effective feature compression and fusion \newline \textbullet~Robustness to noise and interference \newline \textbullet~Unsupervised learning for resource efficiency 
    & \textbullet~High computational overhead for training \newline \textbullet~Sensitivity to hyperparameter tuning \newline \textbullet~Limited Real-World Validation \\ \cline{2-6} 
    & \cite{22_High-precision2023Guo} & MTS-GAN & MTS-GAN is employed for high-precision electromagnetic environment data reconstruction in space-air-ground integrated IoTs. 
    & \textbullet \ High reconstruction accuracy \newline \textbullet~Robustness to severe data loss \newline \textbullet~Adaptability to temporal irregularities
    & \textbullet~Increased training complexity and computational cost \newline \textbullet~Dependence on training data characteristics\\ \cline{2-6} 
    & \cite{a43_Diffusion2024Adam} & GDM & GDM targets interference-aware signal recovery in satellite-terrestrial shared spectrum environments. 
    & \textbullet \ Superior recovery performance \newline \textbullet~Robustness to complex interference \newline \textbullet~Generative capability for high-quality samples
    & \textbullet~Dependence on specific channel conditions \newline \textbullet~High computational overhead for training \newline \textbullet~Large and well-labeled datasets required\\ \cline{2-6} 
    & \cite{25_GS3D2025Adam} & GS3D & GS3D reconstructs signals for Earth stations in motions using RL-guided diffusion and Gaussian splatting. 
    & \textbullet \ Computational efficiency with latent space mapping \newline \textbullet~Superior interference mitigation in dynamic environments \newline \textbullet~No requirement for interference-free training data
    & \textbullet~High computational overhead for training \newline \textbullet~Limited adaptability to extreme dynamic changes \newline \textbullet~Scene scalability limitations \\ \hline
\end{tabular}
\end{table*}

\section{GAI and LLMs for Satellite Network Operation and Remote Sensing Enhancement}\label{sec:application}

It is a critical yet complex goal for modern space systems to optimize satellite network control and system efficiency, as well as to enhance the remote sensing and information acquisition.
However, traditional approaches to network operation and sensing face significant hurdles, including the scale and heterogeneity of satellite constellations, the dynamism of the space environment, and the sheer volume and complexity of sensor data. Advancements in GAI and LLMs offer promising solutions to address these challenges from an agentic AI's perspective. This section aims to explore how GAI and LLMs empower satellite network operations and remote sensing with agentic AI's capabilities.

\subsection{Satellite Network Operations}
Satellite network operations face growing challenges due to increasing constellation density and dynamic mission demands, requiring low-latency, high-efficiency decision-making. Traditional deterministic methods often fail to satisfy the stringent latency and throughput demands of dynamic orbital environments. In contrast, generative techniques, particularly LLMs, provides transformative capabilities by enabling efficient representation learning, context-aware reasoning, and real-time control. Based on these, GAI and LLMs enhance various satellite network operation tasks, such as autonomous spacecraft control \cite{a61_Language2024Victor}, precise orbit prediction \cite{a39_LTE2024Jeong}, accurate behavior inference \cite{a60_Intention2025Heng, a53_Enhanced2020Shen}, semantic communications \cite{a57_On-Air2024Hong-fu}, and access management \cite{a38_A2025Zhen}. These advances demonstrate agentic AI’s potential to support intelligent satellite operations in complex and resource-constrained environments.

\begin{figure}[!t]
    \begin{center}
		\includegraphics[width=1\columnwidth]{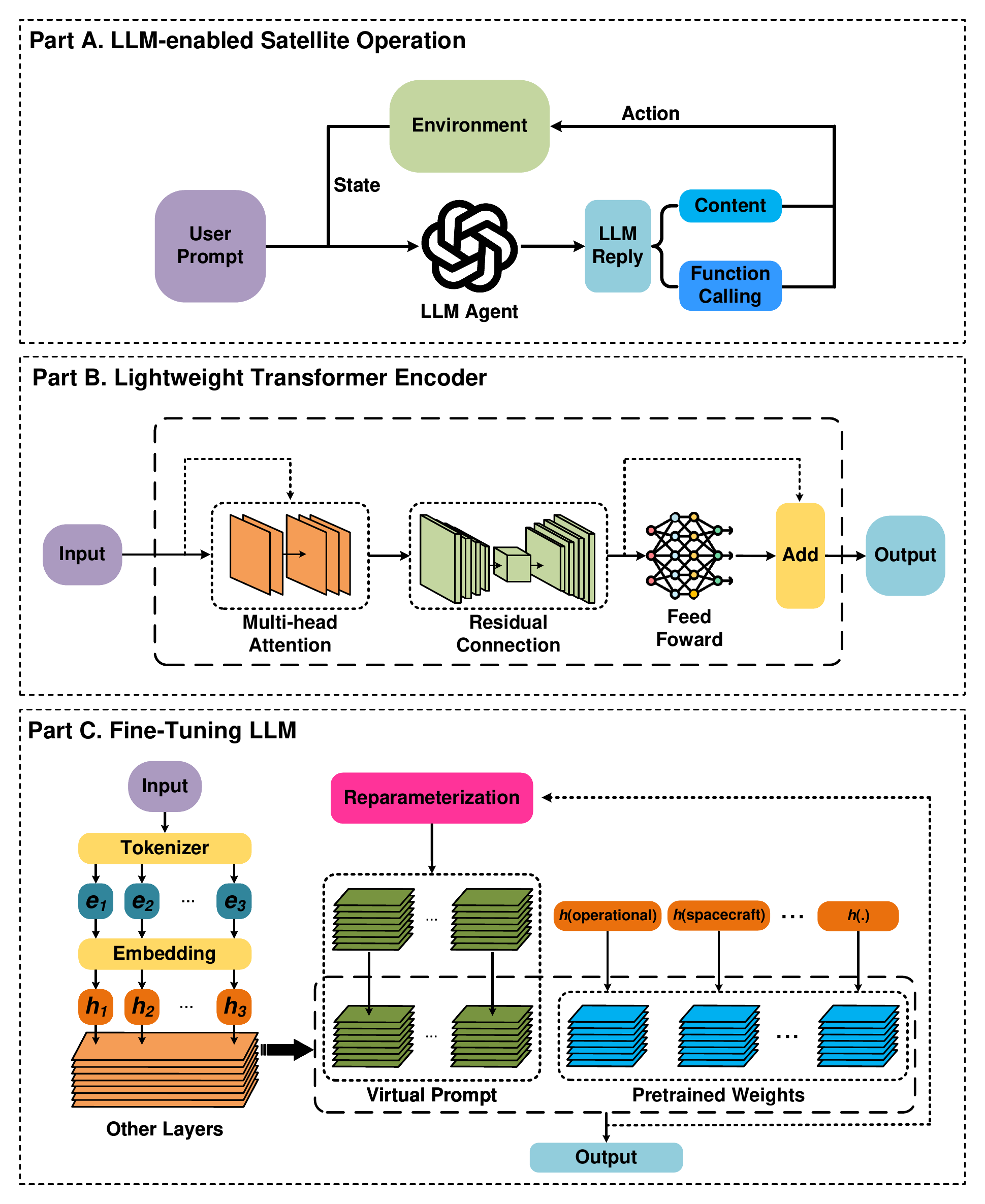}
		\caption{The overview of LLM-based methods for satellite operation and intention recognition. Part A depicts how to use the LLM as an autonomous spacecraft operator in \cite{a61_Language2024Victor}. Based on user prompts and current environmental status, the LLM replies with a reasoned actions expressed as function calling with the specific throttle vector and the textual justification. Part B presents the light weight transformer encoder proposed in \cite{a39_LTE2024Jeong}. The light weight model removes positional encoding and normalization layer to improve prediction performance and reduce computational resources. Part C illustrates the principle of the fine-tuning method in \cite{a60_Intention2025Heng}. This method integrates the newly construct virtual prompt network and the pretrained weights to obtain the fine-tuning model parameter to optimize the output.}
		\label{V-A-1}
	\end{center}
\end{figure}

To support autonomous spacecraft control, the authors in~\cite{a61_Language2024Victor} proposed an LLM-based spacecraft operator for spacecraft guidance, demonstrating the decision-making and reasoning capabilities of agentic AI. With the powerful reasoning and context processing ability of GPT-3.5, the LLM acts as an autonomous spacecraft operator, translating real-time mission telemetry into spacecraft control strategies, as shown in the Part A of Fig.~\ref{V-A-1}. This generative process is achieved through prompt-based reasoning and function calling, enabling the model to generate throttle actions directly from textual orbital state descriptions. In the Kerbal space program differential games (KSPDG), the proposed method achieves a close-to-optimal proximity distance as low as $12.6\,\text{m}$, comparable to expert algorithms like Lambert-MPC. Furthermore, it achieved $0\%$ failure rate using CoT, demonstrating superior sample efficiency and generalization over conventional RL algorithms such as PPO. Impressively, this work highlights the potential of LLMs in autonomous spacecraft operational decision-making.

Beyond control, the growing density of satellites raises the risk of orbital collisions. Accurate orbit prediction is thus critical. While conventional ML approaches often neglect runtime and model size constraints \cite{c38_Satellite2021Puente}, the authors in \cite{a39_LTE2024Jeong} introduced a lightweight transformer encoder for orbit forecasting. As shown in the Part B of  Fig.~\ref{V-A-1}, this proposed model generates future orbital element vectors from multivariate time series inputs, showing the perception capability of agentic AI. By removing positional encoding and layer normalization, it modifies the architecture of transformer encoder, thus enabling efficient generation of accurate satellite orbit predictions through self-attention mechanisms. Trained on $4.8$ million samples from KOMPSAT-3/3A/5, the proposed model outperforms $14$ baselines with MSE reductions of $50.6\%$, $42.4\%$, and $30.0\%$, respectively, and achieves $36.9\%$ faster inference with fewer parameters than the next-best model. This work provides a lightweight and efficient framework for practical orbit tracking.

To enhance collision avoidance, spacecraft also requires to infer the intentions of non-cooperative targets. Existing methods typically rely on motion patterns but overlook interaction dynamics in shared orbits \cite{c39_Self-induced2023Zhang}, which hinders the achievement of optimal avoidance strategies. To address this deficiency, the authors in \cite{a60_Intention2025Heng} proposed an LLM-based framework for intention recognition of space non-cooperative targets, as shown in the Part C of Fig.~\ref{V-A-1}. In this framework, a fine-tuned decision model is used for preliminary behavior classification, while the LLM interprets multi-source task information to refine recognition, demonstrating the contextual reasoning ability of agentic AI. The proposed model predicts a combination of intention and eventually generates concise keywords from the predefined set of intention types. Evaluation across various conditions confirms its superior accuracy and robustness compared to traditional approaches, which indicates LLMs’ aptitude and potential for spacecraft intention recognition tasks under diverse scenarios.

To further advance behavioral understanding in satellite operations, generative models have been explored to synthesize and classify diverse orbital behaviors under uncertainty. In this context, the authors in \cite{a53_Enhanced2020Shen} proposed an improved conditional GAN framework, named space unveiled behavior GAN (SuB-GAN), to support satellite behavior discovery in space situational awareness. By introducing general-sum game theory and Fictitious play into training, the SuB-GAN shows the perception capability of agentic AI to tackle the inherent complexity of orbital dynamics. The proposed model generates realistic orbital behaviors from latent vectors and maneuver labels, enabling both data augmentation and semi-supervised learning. Experiments show that the proposed model achieves $93\%$ classification accuracy, an inception score of $9.7$, and a FID of $0.002$, demonstrating its effectiveness in behavior modeling under uncertainty.

\begin{figure}[!t]
	\begin{center}
		\includegraphics[width=1\columnwidth]{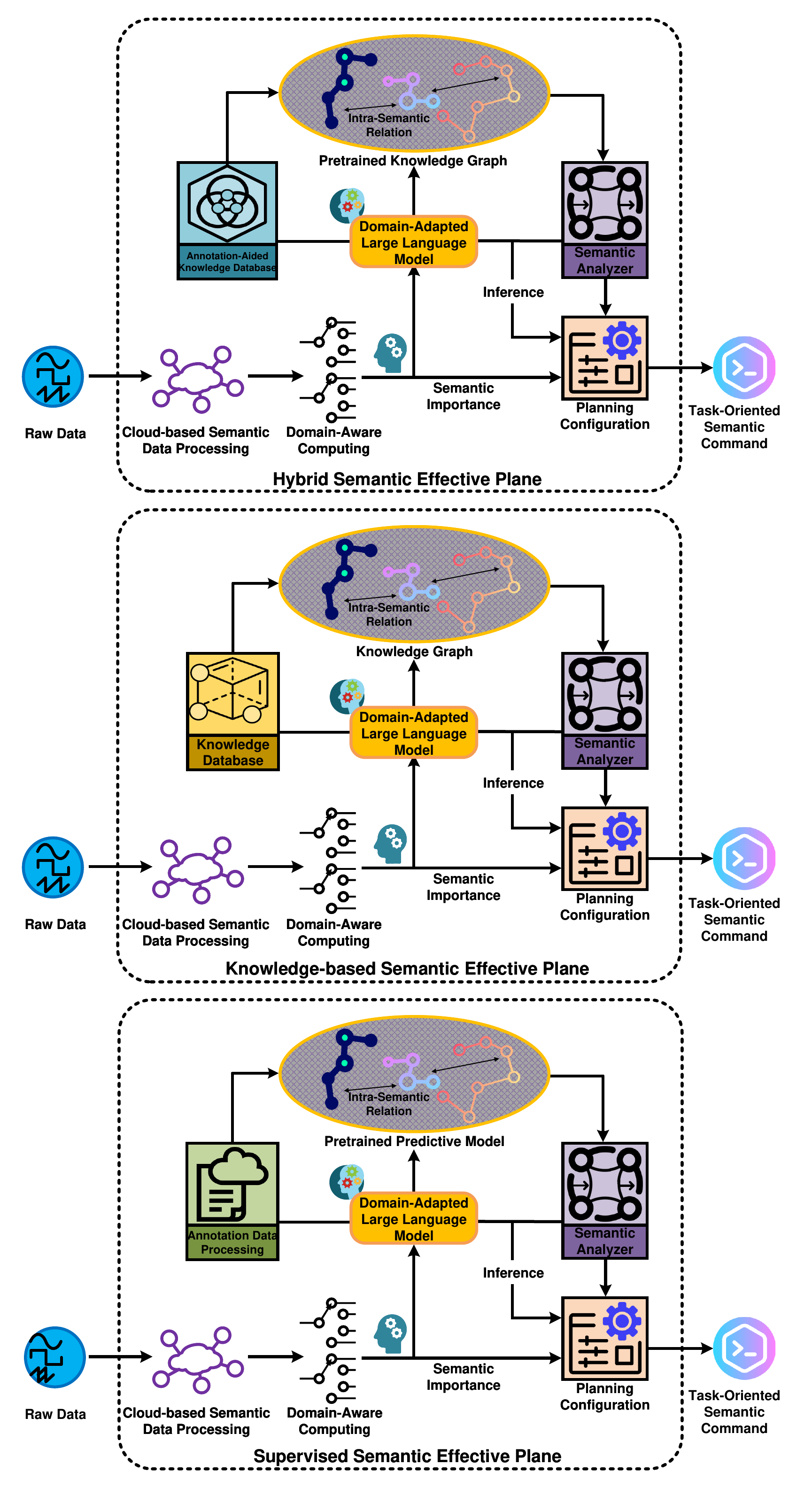}
		\caption{The LLM-based semantic communication framework proposed in \cite{a57_On-Air2024Hong-fu}. By leveraging knowledge graphs, the LLM enriches its understanding of complex tasks through a structured representation of entities and their relationships, allowing for more accurate and context-aware decision-making. The semantic analyzer extracts semantic importance from natural language queries and task descriptions, enhancing the platform’s ability to discern and prioritize critical elements within the text, ensuring that the most relevant information drives the task-solving process. The integration of knowledge graphs and pretrained predictive models enables LLM to refine its task-solving strategies through RL with task feedback, leading to enhanced accuracy and efficiency in addressing complex, real-world scenarios.}
		\label{V-A-2}
	\end{center}
\end{figure}

In parallel, satellite remote sensing systems must manage massive Earth observation data under strict latency constraints. To this end, the authors in \cite{a57_On-Air2024Hong-fu} proposed an LLM-based semantic communication framework for Earth observation system. 
This framework combines domain-adapted LLMs, discrete task-oriented joint source–channel coding (DT-JSCC), and semantic augmentation to extract and optimize task-specific semantic feature vectors derived from the original observational data\cite{c55_Robust2023Xie}, as shown in Fig.~\ref{V-A-2}. The LLM serves as an agentic semantic reasoning module, distilling goal-driven representations aligned with downstream tasks. These optimized vectors, rather than raw data, are transmitted to reduce communication load and energy consumption \cite{c56_Semantics2023Lu}. Furthermore, the vectors are tailored for satellite channel characteristics to enable efficient transmission. Inter-satellite communication and on-board processing are also integrated to further boost efficiency. Simulation results validate that the framework improves classification performance while significantly lowering transmission overhead, highlighting its promise in agentic perception and semantic abstraction for future Earth observation systems.

Moreover, traditional random-access schemes in satellite networks often experience overload under high traffic \cite{c40_NOMA2023Kang}. This degrades scalability and reduces overall throughput. To solve this, the authors in \cite{a38_A2025Zhen} introduced a lightweight TBM for collision detection and load estimation in large-scale satellite-ground vehicle integrated network access scenarios. By incorporating only a two-head attention module and a multi-layer perceptron, the model acquires agentic perceptual capabilities for real-time relevance extraction and rapid decision synthesis from complex traffic patterns. This enables accurate generation of total collision and per-user congestion estimates.
In experiments involving large-scale collisions with up to $300$ UEs and low SNR conditions down to $–14\,\text{dB}$, the proposed scheme achieves a load-estimation accuracy of approximately $100\%$ and a collision-detection probability of around $89\%$, significantly outperforming CNN-based scheme, whose load-estimation accuracy and collision-detection probability fall below $80\%$ and $34\%$, respectively.

\subsection{Satellite Sensing}

\begin{figure*}[!t]
	\begin{center}
		\includegraphics[width=0.8\linewidth]{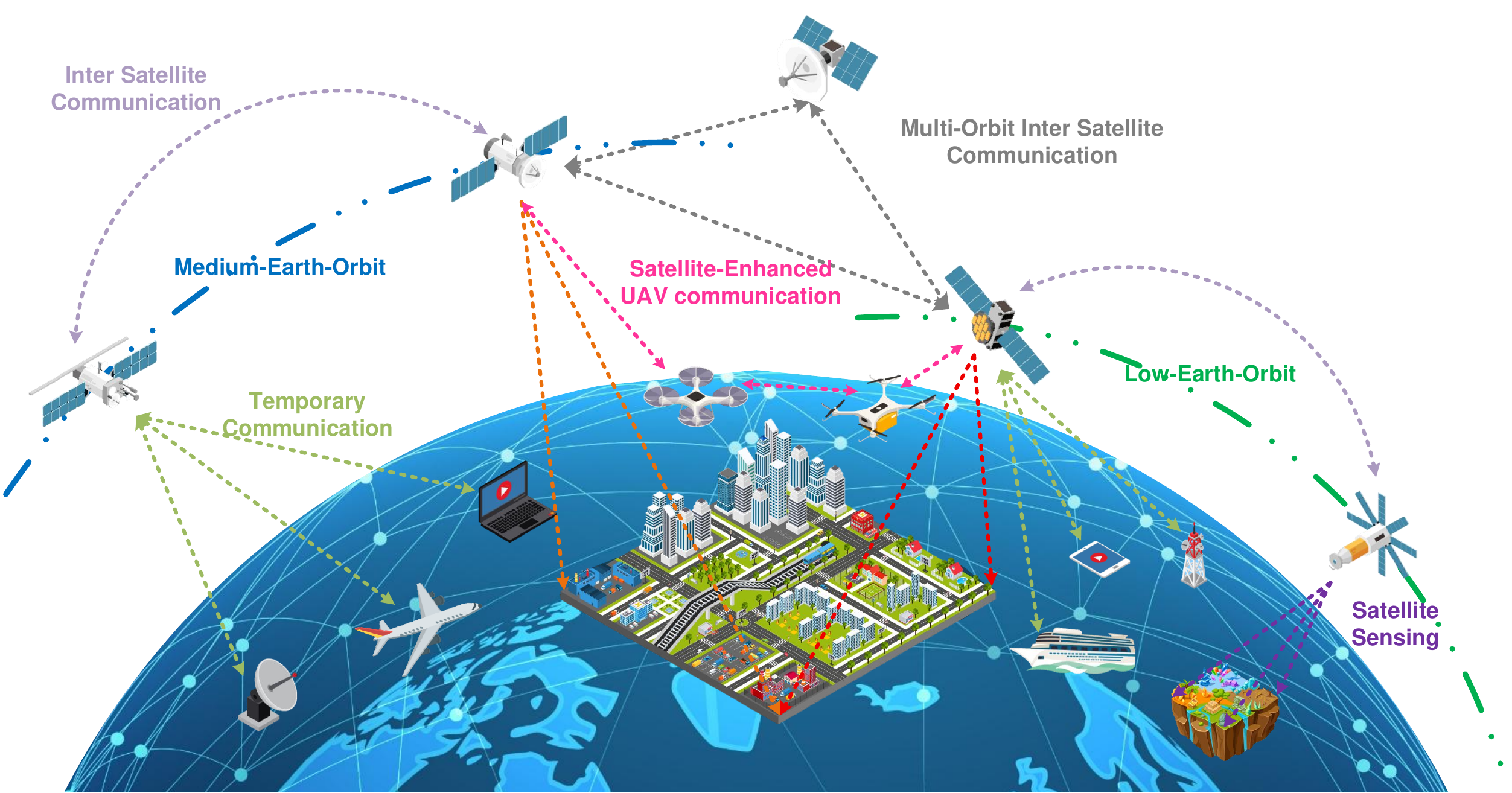}
		\caption{The overall architecture of the satellite communication system model, including temporary communication, satellite-enhanced UAV communication, Earth observation, and co-orbital and multi-orbit inter-satellite communications.}
		\label{V-A-3}
	\end{center}
\end{figure*}

Satellite communication systems support diverse activities, including various forms of temporary communications and satellite sensing, as shown in Fig.~\ref{V-A-3}. In particular, satellite sensing forms a critical foundation for global observation, environmental monitoring, and scientific exploration in SLAETNs. However, inherent challenges such as limited resolution, missing data, and noisy measurements restrict the full utility of existing satellite sensors. Traditional augmentation techniques generally struggle to effectively reconstruct or enhance data acquired under challenging conditions characterized by low quality, high noise, or complex environments. In contrast, GAI has emerged as a transformative solution for real-world applications of satellite sensing, providing the capability for robust data acquisition, analysis, enhancement and multimodal fusion under challenging conditions. Specifically, by leveraging its capability to model intricate data distributions, GAI-based models enable notable improvements in large-scale climate prediction \cite{a54_PrecipGAN2021Wang}, GNSS-based positioning \cite{a35_T-SPP2024Wu}, and satellite image processing\cite{a13_MBGPIN2025Safarov,a46_Super-Resolved2025Ramirez-Jaime,a22_Free2025Chen}. Collectively, these advances underscore agentic AI's significant potential in satellite sensing.

For large-scale climate prediction, the authors in \cite{a54_PrecipGAN2021Wang} proposed PrecipGAN, a conditional GAN framework that merges infrared and sparse microwave inputs to estimate precipitation. The PrecipGAN addresses the challenge of fusing heterogeneous satellite data with limited detection accuracy and coverage  \cite{c41_Evaporation2011Surussavadee}. In the proposed framework, the generator merges infrared and sparse passive microwave inputs to produce high-resolution precipitation maps, while dual discriminators enforce spatial realism and temporal consistency. Through sequence-level adversarial supervision, the PrecipGAN exhibits agentic multi-modal perception and temporal inference, effectively reconstructing precipitation fields beyond direct sensor coverage. Simulation results demonstrate that the proposed PrecipGAN generates more realistic and complete precipitation fields than state-of-the-art baselines like IMERG-Uncal, with $1\,\text{km}$ spatial resolution and smaller error.

In GNSS-based positioning, TBMs have demonstrated strong reasoning capabilities under signal uncertainty. In \cite{a35_T-SPP2024Wu}, the authors proposed D-Tran, a denoising autoencoder transformer model, to enhance GNSS single-point positioning (SPP) in urban canyon environments where multipath and NLoS distortions are common \cite{c42_GNSS2020Zhu}. The proposed D-Tran model combines a transformer for global temporal reasoning and a denoising autoencoder for robust perceptual reconstruction. It produces both estimated corrections and confidence measures, which are fused with traditional SPP outputs. This empowers the system with agentic state estimation and uncertainty awareness. Simulations report a $61\%$ reduction in positioning RMSE and $100\%$ solution availability, significantly outperforming standard SPP in urban canyon scenarios.

In satellite image processing, GAI techniques have shown significant promise in addressing core challenges such as limited resolution, sparse sampling, and signal degradation. These techniques have been customized for distinct data types and sensing objectives, evolving progressively from 2D image enhancement, to 3D light detection and ranging (LiDAR) imaging, and further toward semantic-level recovery in communication scenarios. For super-resolution processing of 2D satellite images, the authors in \cite{a13_MBGPIN2025Safarov} proposed a multi-branch generative prior integration network (MBGPIN). It incorporates multi-scale spatial reasoning through CNN branches and a pretrained vector quantized GAN-based prior stream, enhanced by a hybrid attention mechanism for channel–spatial fusion. This agentic pipeline supports structure-aware generation and high-frequency feature synthesis, enabling realistic image textures while maintaining computational efficiency. Simulation results demonstrate that the MBGPIN outperforms state-of-the-art CNN and transformer-based SR models, achieving higher peak SNR and SSIM than previous methods.

Building upon this, the study of \cite{a46_Super-Resolved2025Ramirez-Jaime} extends the generative framework to 3D terrain modeling using satellite LiDAR data, which overcomes the intrinsic resolution constraints of spaceborne LiDAR systems. In particular, a DDPM-based framework was proposed to enhance the 3D spatial resolution of satellite LiDAR imagery, targeting ecological applications such as forest structure monitoring. This framework generates high-resolution hyperheight data cubes (HHDCs) from sparse measurements \cite{c57_Hyperheight2024Ramirez}. Specifically, it integrates physics-based forward models and side information to align with LiDAR imaging principles \cite{c58_Diffusion2022Chung}. Through iterative denoising, the DDPM demonstrates agentic physical-constrained reconstruction, preserving interpretability in forest monitoring and terrain modeling. The proposed framework was evaluated on forested regions in Florida and Maryland, with a zero-shot test conducted in California. Experimental results show that the DDPM, integrated into the compressive satellite-LiDAR (CS-LiDAR) and concurrent AI spectrometry and adaptive LiDAR system (CASALS), yields enhanced topographic and vegetation imaging, enabling improved forest monitoring and digital terrain modeling.

Furthermore, for the high-robust transmission of perceptual data, semantic communications have been integrated into free-space optical (FSO) \cite{r7_Joint2024Zhang} transmission for remote sensing imagery. To this end, the authors in \cite{a22_Free2025Chen} proposed an FSO-semantic communication scheme based on a VAE-enhanced semantic encoder for large-scale remote sensing image transmission. The model incorporates convolution and attention modules to extract semantically meaningful features and synthesize high-resolution patches. Its agentic encoding supports semantic abstraction and compact generation. Simulation results show that the proposed scheme achieves a $3\,\text{dB}$ power gain, $60\%$ reduction in communication overhead, and comparable image quality, thereby improving both efficiency and resilience.

\textbf{Lesson Learned.} 
As summarized in Table~\ref{GAI in Satellite Network Operation and Remote Sensing}, GAI and LLMs significantly enhance satellite operations across control, transmission, and sensing domains. However, despite their theoretical potential for high performance, these generative methods still face challenges including high computational complexity and reliance on high-quality training data. Future research should focus on balancing computational complexity with performance to enable practical deployment.

\begin{table*}[ht]
\renewcommand\arraystretch{1.2}
\belowrulesep=0pt
\aboverulesep=0pt
\centering
\caption{GAI in Satellite Network Operation and Remote Sensing}
\label{GAI in Satellite Network Operation and Remote Sensing}
\begin{tabular}{>{\centering\arraybackslash}m{0.11\textwidth} | | >{\centering\arraybackslash} m{0.03\textwidth} | >{\centering\arraybackslash}m{0.08\textwidth} | >{\raggedright\arraybackslash} m{0.21\textwidth} | >{\raggedright\arraybackslash}m{0.21\textwidth}| >{\raggedright\arraybackslash} m{0.21\textwidth}}
\toprule
\midrule
    Type  & \multicolumn{1}{c|}{Ref} & Algorithm & Description & Pros & Cons \\ \toprule\midrule
\multirow{6}{*}{
  \centering
  \begin{minipage}[c][6.5cm][c]{\linewidth}
    Satellite Network Operations
  \end{minipage}
}  
    & \cite{a61_Language2024Victor} & LLM & LLM is employed as autonomous agents for spacecraft guidance and control.
    & \textbullet~High precision and adaptability in complex rendezvous tasks \newline  \textbullet~Significant reduction in simulated training runs required
    & \textbullet~Dependence on manual responses to correct the LLM’s actions \newline \textbullet~Dependence on extensive prompt engineering and added telemetry calculations\\ \cline{2-6} 
    
    & \cite{a39_LTE2024Jeong} & Light weight transformer encoder & A modified transformer architecture is employed to enhance the performance of satellite orbit prediction. 
    & \textbullet~Substantial accuracy improvement \newline  \textbullet~Faster execution with marginally fewer parameters 
    & \textbullet~Lack of generalization \newline \textbullet~Limit adaptability to other time-series types data\\ \cline{2-6}
    
    & \cite{a60_Intention2025Heng} & LLM & LLM is employed to improve the accuracy and effectiveness of intention recognition.
    & \textbullet \ High adaptability \newline  \textbullet~High recognition accuracy 
    & \textbullet~Untested real-world robustness and generalization ability \newline \textbullet~High computational demand\\ \cline{2-6}

    & \cite{a53_Enhanced2020Shen} & GAN & GAN is employed to support satellite behavior discovery in space situational awareness.
    & \textbullet~Improved training stability \newline  \textbullet~Semi-supervised capability
    & \textbullet~High computational overhead for training \newline \textbullet~Dependence on initial labeled data \\ \cline{2-6}

    & \cite{a57_On-Air2024Hong-fu} & LLM & LLM-based DT-JSCC framework is employed to improve the communication efficiency.
    & \textbullet~High efficiency and accuracy \newline  \textbullet~Lower overhead and energy consumption
    & \textbullet~Untested real-world robustness and generalization ability \newline \textbullet~High implementation complexity\\ \cline{2-6}

    & \cite{a38_A2025Zhen} & Transformer & A light-weight transformer model is employed to detect collision and estimate load.
    & \textbullet~High accuracy \newline  \textbullet~High computational efficiency 
    & \textbullet~Dependence on predefined channel models \newline \textbullet~High implementation complexity \\
    \hline

\multirow{5}{*}{
  \centering
  \begin{minipage}[c][5cm][c]{\linewidth}
    Satellite Remote Sensing
  \end{minipage}
}

    & \cite{a54_PrecipGAN2021Wang} & PrecipGAN & PrecipGAN is employed to integrates multimodal data in order to achieve accurate satellite-based precipitation estimation.
    & \textbullet \ High resolution \newline \textbullet~Integrates complementary data  
    & \textbullet~High training complexity \newline \textbullet~High computational complexity \\ \cline{2-6} 
    
    & \cite{a35_T-SPP2024Wu} & D-Tran & D-Tran is employed to improve the positioning accuracy.
    & \textbullet~High positioning accuracy \newline \textbullet~Requirement for only single-point GNSS data only
    & \textbullet~High computational complexity \newline \textbullet~Limited generalization \\ \cline{2-6} 
    
    & \cite{a13_MBGPIN2025Safarov} & MBPGIN & MBPGIN is employed to  capture multiscale features and integrate external high-resolution priors. 
    & \textbullet~Superior reconstruction quality \newline  \textbullet~High efficiency and scalability 
    & \textbullet~Limited generalization \newline \textbullet~Artifacts on unseen patterns\\ \cline{2-6} 
    
    & \cite{a46_Super-Resolved2025Ramirez-Jaime} & DDPM & DDPM is employed to achieve super-resolution imaging of satellite LiDAR.
    & \textbullet~High robustness \newline \textbullet~High reconstruction quality 
    & \textbullet~High computational complexity \newline \textbullet~High quality training data required\\  \cline{2-6}

    & \cite{a22_Free2025Chen} & Transformer and VAE & An integrated TBM-VAE architecture is employed to compress images while preserving detail.
    & \textbullet \ Significant link gains \newline \textbullet~Lower overhead 
    & \textbullet~High computational complexity \newline \textbullet~Potential detail loss\\ \hline
    
\end{tabular}
\end{table*}

\section{Future Research Directions}\label{sec:challenges}

\subsection{Dynamic Cloud-Edge-End Collaborative Intelligence for Agentic AI Deployment}
Future research can further explore cloud-edge-device collaboration for the deployment of agentic AI in resource-constrained mobile scenarios to enable low-latency responses, highly reliable services, and optimal global resource allocation \cite{h11_AI-Based2022Wang}. By considering the characteristics of various models and the mobility/resource constraints of different systems, multi-model collaborative frameworks can be designed to efficiently handle complex tasks. For instance, GAI models can be implemented on edge servers to perform data filtering or feature extraction, while complex scheduling process can be conducted on the cloud where LLMs are deployed \cite{h12_Edge2021Cao}. Furthermore, within an agentic AI workflow, GAI models and LLMs can be strategically partitioned across different computational stages. This model partitioning strategy would serve to distribute the computational load more evenly, thereby reducing the resource strain on edge servers and enhancing overall network resource efficiency \cite{h22_Resource-Efficient2024Lai}.

\subsection{Lightweight Distributed Agentic AI for Satellite-Borne and UAV-Borne Scenarios}
Current agentic AI systems, such as those based on GAI and LLMs, encounter energy and computational resource limitations due to the high overhead of the models \cite{s3_Applications2025Vu}. Despite their powerful capabilities, such systems face deployment challenges in edge environments with constrained resources. Future research should focus on developing lightweight, energy-efficient agentic AI frameworks tailored to practical edge scenarios. To this end, strategies, including model pruning, low-rank factorization, and sparse computation methods, need further investigation\cite{h22_Resource-Efficient2024Lai}. For example, estimating gradients of data distributions could replace traditional maximum likelihood training to reduce complexity and energy usage \cite{h23_Efficient2020Pang}. For LLM-based components, techniques such as model distillation and sparse transformers promise to reduce active parameters during inference. Moreover, adaptive parameter adjustment and on-device optimization could be adopted across the full agentic AI pipeline \cite{h24_QIS-GAN2023Zhu}. These techniques promise to yield lightweight models, thus enabling seamless and resource-efficient deployment of edge servers.

\subsection{Cross-Domain Knowledge Transfer and Model Generalization}
Future research can explore data migration that enables efficient transfer of environmental information between the space and aerial layers to overcome sensing distribution disparities and enable cross-platform adaptability \cite{h31_Cooperative2025Zhou}. Additionally, multi-modal generative fusion holds the potential to enhance sensing comprehension and decision robustness. Moreover, by integrating complex scene dynamics and physical constraints as regularization terms, agentic AI methods could further enhance generalization and interpretability \cite{h32_When2024Hadid}. This paves the way for semantically grounded, context-aware decision-making across space-air-ground domains.

\subsection{Hierarchical DT-Driven Simulation for Agentic AI in LAE}
Given the energy-limited and latency-sensitive nature of LAE networks supporting mission-critical services, future research should explore the integration of agentic AI-based generative simulation engines with DT systems. These simulation engines can synthesize realistic environmental and mission dynamics, including airspace traffic, weather, and emergency scenarios, to support real-time decision-making for UAVs and other aerial platforms \cite{h41_Digital2023Souanef}. Within this framework, GAI models enable the creation of diverse synthetic scenarios, while LLMs provide semantic reasoning and instruction generation to support intelligent behavior under uncertainty.
However, conventional DT architectures are typically centralized and not fully aligned with the distributed, cloud-edge-end collaboration model required by agentic AI deployment in dynamic LAE environments. To address this, future work should consider the development of a hierarchical layered DT architecture. In this vision, multi-layer DT systems coordinate cloud-based global models, edge-level situational simulations, and end-device digital agents to collectively enable distributed perception, analysis, and decision-making. Such architecture supports a layer-by-layer realization of simulation and control, allowing for flexible and adaptive support of agentic AI across heterogeneous network infrastructures. This fusion of agentic AI and hierarchical DT will significantly enhance the situational awareness, coordination, and autonomy of LAE systems \cite{h42_LatentFormer2022Amirloo}.

\subsection{Intelligent Autonomous Management of Massive Satellite Networking Enabled LAENets}
The rapid deployment of large-scale LEO constellations introduces significant challenges in orchestrating highly dynamic and mission-critical SLAETNs. This directs future research toward the development of scalable agentic AI frameworks for the autonomous management of massive satellite constellations and their integration with low-altitude networks. These systems must coordinate dynamic topologies, cross-layer routing, and resource allocation across heterogeneous segments. Agentic AI offers a promising solution through multi-agent collaboration, context-aware decision-making, and real-time adaptability. For instance, graph-based models support satellite topology prediction \cite{h51_Combining2025Berreziga}, while LLM components enable intent-driven orchestration \cite{h50_Augmenting2025Roll}. Generative models further aid in scenario simulation and proactive control \cite{h52_Artificial2024AI}. Together, these agentic capabilities support flexible and goal-driven SLAETN management under massive satellite networking.

\subsection{Robustness and Trustworthiness Assurance for Agentic AI in LAE}
Ensuring the robustness and security of agentic AI systems is critical for safety-sensitive LAE applications. Future efforts should strengthen explainability by integrating explainable AI (XAI) techniques with LLMs to provide natural language justifications of AI actions \cite{h61_Explainable2023Dwivedi}. In addition, adversarial defense strategies should be developed to protect against inference manipulation and system destabilization \cite{h62_Resilient2023Wang}. While FL offers privacy by design, communication overheads may arise. Future work can explore parameter compression or privacy-aware coordination mechanisms. Finally, agentic AI systems with generative privacy mechanisms, which combine differential privacy, GAI synthesis, and secure reasoning, could further mitigate risks of sensitive data leakage \cite{h63_An2024Zheng}.

\section{Conclusion}\label{sec:conclusion}
This paper has presented a structured and in-depth survey of generative approaches for SLAETNs, aiming to support the development of agentic AI across integrated space–air–ground infrastructures. We began by introducing the system architecture and unique challenges of SLAETNs, including heterogeneous topologies, dynamic environments, and the need for autonomous operation. Then, a model-centric foundation was established by reviewing five representative classes of generative models, VAEs, GANs, GDMs, TBMs, and LLMs, alongside a comparative analysis of their generative mechanisms, strengths, and constraints in the SLAETN context. Building on this foundation, we investigated how these models contribute to three major functional aspects: communication enhancement, security and privacy reinforcement, and intelligent satellite tasks. Finally, we discussed several future directions for enabling scalable, interpretable, and trustworthy generative agents in SLAETNs. This survey is intended to serve as both a conceptual guide and technical reference for advancing agentic AI in next-generation integrated networks.

\ifCLASSOPTIONcaptionsoff
  \newpage
\fi

\bibliographystyle{IEEEtran}
\bibliography{main}

\end{document}